\documentclass[journal,transmag]{IEEEtran}
\usepackage[subfigure]{graphfig}
\usepackage{graphicx}
\usepackage{amsmath,amsfonts,amssymb}
\usepackage{bm}
\usepackage{mathrsfs}
\newtheorem{theorem}{Theorem}[section]
\newtheorem{lemma}[theorem]{Lemma}

\newtheorem{corollary}[theorem]{Corollary}

\newtheorem{definition}[theorem]{Definition}

\ifCLASSINFOpdf
\else
\fi
\begin{document}
%
\title{The convergence guarantee of the iterative thresholding algorithm with suboptimal feedbacks for large systems}


\author{\IEEEauthorblockN{Zhanjie Song\IEEEauthorrefmark{1},
Shidong Li\IEEEauthorrefmark{2},
Ningning Han\IEEEauthorrefmark{1},
}
\IEEEauthorblockA{\IEEEauthorrefmark{1}School of mathematics, Tianjin University, Tianjin 300354, China}
\IEEEauthorblockA{\IEEEauthorrefmark{2}Department of Mathematics, San Francisco State University, San Francisco, CA94132,USA}

\thanks{Manuscript received December 1, 2012; revised August 26, 2015.
Corresponding author: Shidong Li (email:shidong@sfsu.edu).}}

\markboth{$\times\times\times\times\times$}%
{Shell \MakeLowercase{\textit{et al.}}: Bare Demo of IEEEtran.cls for IEEE Transactions on Magnetics Journals}
%



\IEEEtitleabstractindextext{%
\begin{abstract}
 Thresholding based iterative algorithms have the trade-off between effectiveness and optimality.  Some are effective but involving sub-matrix inversions in every step of iterations.  For systems of large sizes, such algorithms can be computationally expensive and/or prohibitive. The null space tuning algorithm with hard thresholding and feedbacks (NST+HT+FB) has a mean to expedite its procedure by a suboptimal feedback, in which sub-matrix inversion is replaced by an eigenvalue-based approximation.  The resulting suboptimal feedback scheme becomes exceedingly effective for large system recovery problems. An adaptive algorithm based on thresholding, suboptimal feedback and null space tuning (AdptNST+HT+subOptFB) without a prior knowledge of the sparsity level is also proposed and analyzed. Convergence analysis is the focus of this article.  Numerical simulations are also carried out to demonstrate the superior efficiency of the algorithm compared with state-of-the-art iterative thresholding algorithms at the same level of recovery accuracy, particularly for large systems.
\end{abstract}

}

\maketitle

\IEEEdisplaynontitleabstractindextext

%
\IEEEpeerreviewmaketitle

\section{Introduction}
%
%
%
%
\IEEEPARstart{T}{he} problem of compressive sensing (CS) is to recover sparse signals from incomplete linear measurements
\begin{equation}\label{eq1}
y=Ax,
\end{equation}
where $A\in\mathbb{R}^{M\times N}$ is the sampling matrix with $M\ll N$, and $x$ is the $N$-dimensional sparse signal with only $s\ll N $ nonzero coefficients.

Since most of natural signals are sparse or highly compressible under a basis, CS has a wide range of applications including image processing \cite{yang2010image}, radar detection and estimation \cite{duarte2013spectral}, source localization with sensor arrays \cite{haupt2008compressed}, biological application \cite{parvaresh2008recovering}, sub-Nyquist sampling \cite{gedalyahu2010time}, data separation \cite{song} etc. Various algorithms have been proposed for solving problem (\ref{eq1}). Evidently, the underlying model involves finding the sparsest solutions satisfying the linear equation,
\begin{equation}\label{eq2}
\min \limits_{x\in\mathbb{R}^{N}}\|x\|_{0},\ \ s.t.\ \ y=Ax,
\end{equation}
or, in its Lagrangian version
\begin{equation}\label{eq3}
\min \limits_{x\in\mathbb{R}^{N}}\|y-Ax\|_{2}^{2}+\lambda\|x\|_{0},
\end{equation}
where $\|x\|_{0}$ is the measurement of the number of nonzero entries in $x$ and $\lambda$ is a regularization parameter.
(\ref{eq2}) and (\ref{eq3}) are clearly combinatorial and computationally intractable \cite{natarajan1995sparse}.

Heuristic (tractable) algorithms have been extensively studied to solve this problem. A common strategy is based on problem relaxation that replaces the $\ell_{0}$-norm by an $\ell_{p}$-norm (with $0<p\leq1$) \cite{lisong}. The relaxed problem can be solved efficiently by simple optimization procedures. Well-known algorithms based on such an strategy are basis pursuit \cite{chen2001atomic}, least absolute shrinkage and selection operator \cite{tibshirani1996regression}, focal underdetermined system solver \cite{gorodnitsky1997sparse}, bregman iterations \cite{cai2009linearized}, \cite{yin2008bregman} algorithms. The pursuit algorithms are another class of popular approaches, which build up the sparse solutions by making a series of greedy decisions. Typical representative approaches are matching pursuit \cite{mallat1993matching}, orthogonal matching pursuit \cite{pati1993orthogonal}. A number of improved greedy pursuit algorithms have also been put forward, e.g., stagewise orthogonal matching pursuit \cite{donoho2012sparse}, compressive sampling matching pursuit \cite{needell2009cosamp} and subspace pursuit \cite{dai2009subspace}, etc.
A popular category for CS is the iterative thresholding/shrinkage algorithm, which has attracted much attention due to its remarkable performance/complexity trade-off. These methods recover the sparse signal by making a succession of thresholding operations. The iterative hard thresholding algorithm was first introduced by Blumensath and Davies in \cite{blumensath2008iterative},\cite{blumensath2009iterative}. Following a similar procedure, in \cite{daubechies2004iterative}, Daubechies et al proposed  a soft thresholding operation by replacing the $\ell_{0}$-norm by the $\ell_{1}$-norm. The convergence analysis of soft thresholding algorithm were also shown in \cite{bredies2008linear}. In \cite{maleki2010optimally}, Donoho and Maliki combined an exact solution to a small linear system with thresholding before and after the solution to derive a more sophisticated scheme, named two-stage thresholding method. Very recently, Foucart proposed a hard thresholding pursuit algorithm \cite{foucart2011hard} and a graded hard thresholding pursuit algorithm \cite{bouchot2016hard}. In fact, the hard thresholding pursuit algorithm can be regarded as a hybrid of the iterative hard thresholding algorithm and the compressive sampling matching pursuit.

Although iterative thresholding algorithms provide good performance, the matrix inversion among these algorithms is computationally prohibitive for large-scale data. In \cite{li2014fast}, the null space tuning algorithm with hard-thresholding and feedbacks (NST+HT+FB) was proposed to find sparse solutions. The convergence results about NST+HT+FB was presented in previous articles \cite{li2014fast}.

The aim of this article is to study the theoretical convergence of a further improved adaptive suboptimal feedback scheme AdptNST+HT+subOptFB, assuming no knowledge of the sparsity level.  In a simpler case, if the sparsity level is known, one can obtain the convergence guarantee of a suboptimal scheme NST+HT+subOptFB.

For clarity, the following are some of the notations used in this article. $S$ is the true support of the $s$-sparse vector $x$. $x_{T}$ is the restriction of a vector $x\in\mathbb{R}^{N}$ to an index set $T$, by $T^{c}$ the complement set of $T$ in $\{1, 2, . . . , N\}$, and by $A_{T}$ the sub-matrix consisting of columns of $A$ indexed by $T$, respectively. $|T|$ is the cardinality of set $T$. $T\triangle T'$ is the symmetric difference of $T$ and $T'$, i.e., $T\triangle T' =(T\setminus T')\cup(T'\setminus T)$.

\section{Preliminary results}
This section gives some previous arguments and collects the key lemmas needed in the converge analysis of AdptNST+HT+subOptFB.


\subsection{Problem Statements}
The iterative framework of the approximation and null space tuning (NST) algorithms is as follows
\begin{equation*} \label{eq4}
\text{(NST)}\ \ \ \
\left\{ \begin{aligned}
         \begin{aligned}
         &\mu^{k}=\mathbb{D}(x^{k}),\\
         &x^{k+1}=x^{k}+\mathbb{P}(\mu^{k}-x^{k}).\\
         \end{aligned}
         \end{aligned} \right.
\end{equation*}
Here $\mathbb{D}(x^{k})$ approximates the desired solution by various principles, and $\mathbb{P}:=I-A^{\ast}(AA^{\ast})^{-1}A$ is the orthogonal projection
onto ker$A$. The feasibility of $x^{0}$ is assumed, which guarantees that the sequence $\{x^{k}\}$ are all feasible. Obviously, $\mu^{k}\rightarrow x$ is
expected as $k$ increases. Since the sequence $\{x^{k}\}$ are always feasible in the framework of the NST algorithms, one may split $y$ as
\begin{equation*}\label{eq5}
y=Ax=A_{T_{k}}x_{T_{k}}^{k}+A_{T^{c}_{k}}x_{T^{c}_{k}}^{k}.
\end{equation*}
In most (if not all) thresholding algorithms, thresholding (hard or soft) is taken by merely keeping the entries of $x^{k}$ on $T_{k}$,
and completely abandons the contribution of $A_{T^{c}_{k}}x_{T^{c}_{k}}^{k}$ to the measurement $y$. Though $x_{T^{c}_{k}}^{k}$ gradually diminishes as $k\rightarrow\infty$, it is not difficult to observe that the contribution of $A_{T^{c}_{k}}x_{T^{c}_{k}}^{k}$ to $y$ can be quite significant at initial
iterations. Therefore, simple thresholding alone can be quite infeasible at earlier stages. The mechanism of feedback is to feed the contribution of $A_{T_{k}^{c}}x_{T_{k}^{c}}^{k}$ to $y$ back to im($A_{T_{k}}$), the image of $A_{T_{k}}$. A straightforward way is to set
\begin{equation*}\label{eq6}
\eta^{k}=\arg\min \limits_{\eta}\|A_{T_{k}}\eta-A_{T_{k}^{c}}x_{T_{k}^{c}}^{k}\|_{2},
\end{equation*}
which has the best/least-square solution
\begin{equation*}\label{eq7}
\eta^{k}=(A_{T_{k}}^{\ast}A_{T_{k}})^{-1}A_{T_{k}}^{\ast}A_{T_{k}^{c}}x_{T_{k}^{c}}^{k}.
\end{equation*}
The NST+HT+FB algorithm is then established as follows
\begin{equation*} \label{eq8}
\text{(NST+HT+FB)}\ \ \ \
\left\{ \begin{aligned}
         \begin{aligned}
         &\mu_{T_{k}}^{k}=x_{T_{k}}^{k}+(A_{T_{k}}^{\ast}A_{T_{k}})^{-1}A_{T_{k}}^{\ast}A_{T^{c}_{k}}x_{T^{c}_{k}}^{k}, \\
         &\mu_{T^{c}_{k}}^{k}=0, \\
         &x^{k+1}=x^{k}+\mathbb{P}(\mu^{k}-x^{k}),\\
         \end{aligned}
         \end{aligned} \right.
\end{equation*}
where $\mathbb{P}=I-A^{\ast}(AA^{\ast})^{-1}A$ and $|T_{k}|=s$. The null space tuning (NST) step can be rewritten as $x^{k+1}=\mu^{k}+A^{\ast}(AA^{\ast})^{-1}(y-A\mu^{k})$.

The role of $(A_{T_{k}}^{\ast}A_{T_{k}})^{-1}A_{T_{k}}^{\ast}A_{T^{c}_{k}}x_{T^{c}_{k}}^{k}$ is to calculate the feedback (to the index set $T_k$) of the tail contribution to $y$. One can see that the matrix inversion can also be expensive and/or computationally prohibitive for large-scale data.

However, the feedback mechanism seems to be exceedingly convenient to derive a much less expensive suboptimal feedback scheme.  In fact, as long as $A_{T_{k}}^{\ast}A_{T_{k}}$ is well conditioned, which is a typical requirement by the well-known restricted isometry property (RIP), $(A_{T_{k}}^{\ast}A_{T_{k}})^{-1}$ can be approximated by $\lambda^{-1} I$ with $\lambda^{-1}$ being on the order of the spectrum of $(A_{T_{k}}^{\ast}A_{T_{k}})^{-1}$. Evidently, a natural approximation of the feedback can be simplified to $\lambda^{-1} A_{T_{k}}^{\ast}A_{T^{c}_{k}}x_{T^{c}_{k}}^{k}$. We then reach the suboptimal feedback scheme
(NST+HT+subOptFB)
\begin{equation*} \label{eq9}
\text{(NST+HT+subOptFB)}\ \ \ \
\left\{ \begin{aligned}
         \begin{aligned}
         &\mu_{T_{k}}^{k}=x_{T_{k}}^{k}+\lambda^{-1}A_{T_{k}}^{\ast}A_{T^{c}_{k}}x_{T^{c}_{k}}^{k}, \\
         &\mu_{T^{c}_{k}}^{k}=0, \\
         &x^{k+1}=x^{k}+\mathbb{P}(\mu^{k}-x^{k}).\\
         \end{aligned}
         \end{aligned} \right.
\end{equation*}
The typical NST+HT+subOptFB scheme, like in all other non-adaptive approaches, requires a prior estimation or knowledge of the sparsity $s$, and $|T_{k}|=s$ at each iteration.  A further improved adaptive suboptimal feedback algorithm (AdptNST+HT+subOptFB), is also proposed and studied in this article.  In the AdptNST+HT+subOptFB scheme, the size of the index set increases with the iteration. Precisely, a sequence $\{\mu^{k}\}$ of $k$-sparse vectors is constructed according to
 \begin{equation} \label{eq9}
\text{(AdptNST+HT+subOptFB)}\ \ \ \
\left\{ \begin{aligned}
         \begin{aligned}
         &\mu_{T_{k}}^{k}=x_{T_{k}}^{k}+\lambda^{-1}A_{T_{k}}^{\ast}A_{T^{c}_{k}}x_{T^{c}_{k}}^{k}, \\
         &\mu_{T^{c}_{k}}^{k}=0, \\
         &x^{k+1}=x^{k}+\mathbb{P}(\mu^{k}-x^{k}),\\
         \end{aligned}
         \end{aligned} \right.
\end{equation}
where $|T_{k}|=k$ at each iteration, clearly increasing along the number of iteration $k$. The convergence result is also provided. It shows that under a preconditioned restricted isometry constant of the matrix $A$ and the constant $\lambda$, AdptNST+HT+subFB indeed obtains all $s$-sparse signals.

\subsection{Properties characterized via RIP (P-RIP)}
\begin{definition} \label{de1}
\cite{candes2005decoding} For each integer $s = 1, 2,... ,$ the restricted isometry constant $\delta_{s}$ of a matrix $A$ is defined as the smallest number $\delta_{s}$ such that
\begin{equation*}\label{eq10}
(1-\delta_{s})\|x\|_{2}^{2}\leq\|Ax\|_{2}^{2}\leq(1+\delta_{s})\|x\|_{2}^{2}
\end{equation*}
holds for all $s$-sparse vectors $x$. Equivalently, it can be given by
\begin{equation*}\label{eq13}
\delta_{s}=\max \limits_{S\subset[N],|S|\leq s}\|A_{S}^{\ast}A_{S}-I\|_{2}.
\end{equation*}
\end{definition}

\begin{definition} \label{de2}
 \cite{li2014fast}. For each integer $s = 1, 2,... ,$ the preconditioned restricted isometry constant $\gamma_{s}$ of a matrix $A$ is defined as the smallest number $\gamma_{s}$ such that
\begin{equation*}\label{eq11}
(1-\gamma_{s})\|x\|_{2}^{2}\leq\|(AA^{\ast})^{-\frac{1}{2}}Ax\|_{2}^{2}
\end{equation*}
holds for all $s$-sparse vectors $x$.
In fact, the preconditioned restricted isometry constant $\gamma_{s}$ characterizes the restricted isometry property of the preconditioned matrix $(AA^{\ast})^{-\frac{1}{2}}A$. Since
\begin{equation*}
\|(AA^{\ast})^{-\frac{1}{2}}Ax\|_{2}\leq\|(AA^{\ast})^{-\frac{1}{2}}A\|_{2}\|x\|_{2}=\|x\|_{2},
\end{equation*}
$\gamma_{s}$ is actually the smallest number such that, for all $s$-sparse vectors $x$,
\begin{equation*}\label{eq12}
(1-\gamma_{s})\|x\|_{2}^{2}\leq\|(AA^{\ast})^{-\frac{1}{2}}Ax\|_{2}^{2}\leq(1+\gamma_{s})\|x\|_{2}^{2}.
\end{equation*}
Note that $\gamma_{s}(A)=\delta_{s}((AA^{\ast})^{-\frac{1}{2}}A)$. Evidently, for Parseval frames, since $AA^{\ast}=I$, $\gamma_{s}(A)=\delta_{s}(A)$.
Equivalently, it can also be given by
\begin{equation*}\label{eq14}
\gamma_{s}=\max \limits_{S\subset[N],|S|\leq s}\|A_{S}^{\ast}(AA^\ast)^{-1}A_{S}-I\|_{2}.
\end{equation*}
\end{definition}

\begin{lemma} \label{le1}
 Let $u,v\in\mathbb{R}^{N}$ be vectors with supp$(u)=T'$ and supp$(v)=T''$. If $|T'\cup T''|\leq t$, then
\begin{equation}\label{eq15}
|\langle u,(I-A^{\ast}(AA^\ast)^{-1}A)v\rangle|\leq\gamma_{t}\|u\|_{2}\|v\|_{2},
\end{equation}
\begin{equation}\label{eq16}
\|[(I-A^{\ast}(AA^\ast)^{-1}A)v]_{T'}\|_{2}\leq\gamma_{t}\|v\|_{2}.
\end{equation}
\end{lemma}

$\bm{Proof.}$
Indeed, setting $T=T'\cup T''$, one has
\begin{equation*}
\begin{aligned}
&|\langle u,(I-A^{\ast}(AA^\ast)^{-1}A)v\rangle|\\
&=|\langle u,v\rangle-\langle Au,(AA^\ast)^{-1}Av\rangle|\\
&=|\langle u_{T},v_{T}\rangle-\langle A_{T}u_{T},(AA^\ast)^{-1}A_{T}v_{T}\rangle|\\
&=|\langle u_{T},(I-A_{T}^{\ast}(AA^\ast)^{-1}A_{T})v_{T}\rangle|\\
&\leq\|u_{T}\|_{2}\|(I-A_{T}^{\ast}(AA^\ast)^{-1}A_{T})v_{T}\|_{2}\\
&\leq\|u_{T}\|_{2}\|(I-A_{T}^{\ast}(AA^\ast)^{-1}A_{T})\|_{2}\|v_{T}\|_{2}\\
&\leq\|u_{T}\|_{2}\gamma_{t}\|v_{T}\|_{2},\\
\end{aligned}
\end{equation*}
where the last step is based on {\em Definition \ref{de2}}. It is easy to observe that
\begin{equation*}
|\langle u,(I-A^{\ast}(AA^\ast)^{-1}A)v\rangle|\leq\gamma_{t}\|u\|_{2}\|v\|_{2},
\end{equation*}
Using (\ref{eq15}), we have
\begin{equation*}
\begin{aligned}
&\|[(I-A^{\ast}(AA^\ast)^{-1}A)v]_{T'}\|_{2}^{2}\\
&=\langle[(I-A^{\ast}(AA^\ast)^{-1}A)v]_{T'},(I-A^{\ast}(AA^\ast)^{-1}A)v\rangle\\
&\leq\gamma_{t}\|[(I-A^{\ast}(AA^\ast)^{-1}A)v]_{T'}\|_{2}\|v\|_{2},
\end{aligned}
\end{equation*}
and it remains to simplify by solving for $\|[(I-A^{\ast}(AA^\ast)^{-1}A)v]_{T'}\|_{2}$ to obtain (\ref{eq16}).

\begin{lemma} \label{le2}
 Let $u,v\in\mathbb{R}^{N}$ be vectors with supp$(u)=T'$ and supp$(v)=T''$. If $|T'|\leq s$, $|T''|\leq t$ and $T'\cap T''=\emptyset$, then
\begin{equation}\label{add}
|\langle Au,(AA^\ast)^{-1}Av\rangle|\leq\gamma_{s+t}\|u\|_{2}\|v\|_{2},
\end{equation}
\begin{equation*}
\|A^{\ast}_{T'}(AA^\ast)^{-1}A_{T''}v_{T''}\|_{2}\leq\gamma_{s+t}\|v\|_{2}.
\end{equation*}
\end{lemma}

$\bm{Proof.}$
 Let $T=T'\cup T''$, note that $T'\cap T''=\emptyset$, we have $\langle u_{T},v_{T}\rangle=0$.
\begin{equation*}
\begin{aligned}
&|\langle Au,(AA^\ast)^{-1}Av\rangle|\\
&=|\langle A_{T}u_{T},(AA^\ast)^{-1}A_{T}v_{T}\rangle|\\
&=|\langle A_{T}u_{T},(AA^\ast)^{-1}A_{T}v_{T}\rangle-\langle u_{T},v_{T}\rangle|\\
&=|\langle(A_{T}^{\ast}(AA^\ast)^{-1}A_{T}-I)v_{T},u_{T}\rangle|\\
&\leq\|(A_{T}^{\ast}(AA^\ast)^{-1}A_{T}-I)v_{T}\|_{2}\|u_{T}\|_{2}\\
&\leq\|A_{T}^{\ast}(AA^\ast)^{-1}A_{T}-I\|_{2}\|v_{T}\|_{2}\|u_{T}\|_{2}\\
&\leq\gamma_{s+t}\|u\|_{2}\|v\|_{2}.
\end{aligned}
\end{equation*}

\begin{equation*}
\begin{aligned}
&\|A^{\ast}_{T'}(AA^\ast)^{-1}A_{T''}v_{T''}\|_{2}^{2}\\
&=|\langle A^{\ast}_{T'}(AA^\ast)^{-1}A_{T''}v_{T''},A^{\ast}_{T'}(AA^\ast)^{-1}A_{T''}v_{T''}\rangle|\\
&=|\langle (AA^\ast)^{-1}A_{T''}v_{T''},A_{T'}A^{\ast}_{T'}(AA^\ast)^{-1}A_{T''}v_{T''}\rangle|\\
&=|\langle (AA^\ast)^{-1}Av,A_{T'}A^{\ast}_{T'}(AA^\ast)^{-1}A_{T''}v_{T''}\rangle|\\
&\leq\gamma_{s+t}\|v\|_{2}\|A^{\ast}_{T'}(AA^\ast)^{-1}A_{T''}v_{T''}\|_{2},\\
\end{aligned}
\end{equation*}
where the last step is due to (\ref{add}).
Therefore,
$\|A^{\ast}_{T'}(AA^\ast)^{-1}A_{T''}v_{T''}\|_{2}\leq\gamma_{s+t}\|v\|_{2}.$

\begin{lemma} \label{le3}
 For any $e\in\mathbb{R}^{M}$ and $|T|\leq t$,
\begin{equation*}
\|(A^{\ast}(AA^{\ast})^{-1}e)_{T}\|_{2}\leq\sqrt{1+\theta_{t}}\|e\|_{2},
\end{equation*}
where $\theta_{t}(A)=\delta_{t}((AA^\ast)^{-1}A)$.
\end{lemma}

$\bm{Proof.}$
 Using the fact that
\begin{equation*}
\begin{aligned}
\|(A^{\ast}(AA^{\ast})^{-1}e)_{T}\|_{2}^{2}&=\langle(A^{\ast}(AA^{\ast})^{-1}e)_{T},A^{\ast}(AA^{\ast})^{-1}e\rangle\\
&=\langle e,(AA^{\ast})^{-1}A(A^{\ast}(AA^{\ast})^{-1}e)_{T}\rangle\\
&\leq\|e\|_{2}\|(AA^{\ast})^{-1}A(A^{\ast}(AA^{\ast})^{-1}e)_{T}\|_{2}\\
&=\|e\|_{2}\sqrt{1+\theta_{t}}\|(A^{\ast}(AA^{\ast})^{-1}e)_{T}\|_{2},
\end{aligned}
\end{equation*}
we have
$\|(A^{\ast}(AA^{\ast})^{-1}e)_{T}\|_{2}\leq\sqrt{1+\theta_{t}}\|e\|_{2}.$

\section{The convergence analysis of AdptNST+HT+subOptFB}
In this section, we derive the main result of this paper. The result shows that AdptNST+HT+subOptFB are guaranteed to converge under a P-RIP condition of $A$ and the value of $\lambda$.
\subsection{The main result}
The AdptNST+HT+subOptFB generates a sequence $\{\mu^{k}\}$ of $k$-sparse vectors according to (\ref{eq9}) with $|T_{k}|=k$ at each iteration. We can first obtain the following argument based on NST, i.e., $x^{k+1}=\mu^{k}+A^{\ast}(AA^{\ast})^{-1}(y-A\mu^{k})$.

\begin{lemma} \label{le4}
 Suppose $y=Ax+e$ where $x\in\mathbb{R}^{N}$ is $s$-sparse with $S=$supp$(x)$ and $e\in\mathbb{R}^{M}$ is the measurement error. If $\mu'\in\mathbb{R}^{N}$ is $s'$-sparse and $T$ is an index set of $t\geq s$ largest absolute entries of $\mu'+A^{\ast}(AA^\ast)^{-1}(y-A\mu')$, then
\begin{equation}\label{eq17}
\|x_{T^{c}}\|\leq\sqrt{2}(\gamma_{s+s'+t}\|x-\mu'\|_{2}+\sqrt{1+\theta_{t+s}}\|e\|_{2}),
\end{equation}
where $\theta_{s}(A)=\delta_{s}((AA^\ast)^{-1}A)$.
\end{lemma}

$\bm{Proof.}$
 By the above definition, we have that
\begin{equation*}
\begin{aligned}
&\|[\mu'+A^{\ast}(AA^\ast)^{-1}(y-A\mu')]_{T}\|_{2}^{2}\\
&\geq\|[\mu'+A^{\ast}(AA^\ast)^{-1}(y-A\mu')]_{S}\|_{2}^{2}.\\
\end{aligned}
\end{equation*}

Eliminating the common terms over $T\bigcap S$, one has
\begin{equation*}
\begin{aligned}
&\|[\mu'+A^{\ast}(AA^\ast)^{-1}(y-A\mu')]_{T\setminus S}\|_{2}\\
&\geq\|[\mu'+A^{\ast}(AA^\ast)^{-1}(y-A\mu')]_{S\setminus T}\|_{2}.\\
\end{aligned}
\end{equation*}

For the left hand,
\begin{equation*}
\begin{aligned}
&\|[\mu'+A^{\ast}(AA^\ast)^{-1}(y-A\mu')]_{T\setminus S}\|_{2}\\
&=\|[\mu'-x+A^{\ast}(AA^\ast)^{-1}(Ax+e-A\mu')]_{T\setminus S}\|_{2}\\
&=\|[(I-A^{\ast}(AA^\ast)^{-1}A)(\mu'-x)+A^{\ast}(AA^\ast)^{-1}e]_{T\setminus S}\|_{2}.\\
\end{aligned}
\end{equation*}

The right hand satisfies
\begin{equation*}
\begin{aligned}
&\|[\mu'+A^{\ast}(AA^\ast)^{-1}(y-A\mu')]_{S\setminus T}\|_{2}\\
&=\|[\mu'+A^{\ast}(AA^\ast)^{-1}(Ax+e-A\mu')+x-x]_{S\setminus T}\|_{2}\\
&\geq\|x_{S\setminus T}\|_{2}-\|[(I-A^{\ast}(AA^\ast)^{-1}A)(\mu'-x)+A^{\ast}(AA^\ast)^{-1}e]_{S\setminus T}\|_{2}.\\
\end{aligned}
\end{equation*}

Therefore, we obtain
\begin{equation*}
\begin{aligned}
&\|x_{S\setminus T}\|_{2}\\
&\leq \|[(I-A^{\ast}(AA^\ast)^{-1}A)(\mu'-x)+A^{\ast}(AA^\ast)^{-1}e]_{S\setminus T}\|_{2}\\
&+\|[(I-A^{\ast}(AA^\ast)^{-1}A)(\mu'-x)_{T\setminus S}+A^{\ast}(AA^\ast)^{-1}e]_{T\setminus S}\|_{2}\\
&\leq\sqrt{2}\|[(I-A^{\ast}(AA^\ast)^{-1}A)(\mu'-x)+A^{\ast}(AA^\ast)^{-1}e]_{T\triangle S}\|_{2}\\
&\leq\sqrt{2}\|[(I-A^{\ast}(AA^\ast)^{-1}A)(\mu'-x)]_{T\triangle S}\|_{2}\\
&+\sqrt{2}\|[A^{\ast}(AA^\ast)^{-1}e]_{T\triangle S}\|_{2}\\
&\leq\sqrt{2}(\gamma_{s+s'+t}\|x-\mu'\|_{2}+\sqrt{1+\theta_{t+s}}\|e\|_{2}).
\end{aligned}
\end{equation*}
where the last step uses {\em Lemma \ref{le1}} and {\em Lemma \ref{le3}}.

The main contribution of this paper can be formally stated as the following theorem.
\begin{theorem} \label{the1}
 Suppose that the P-RIP of the matrix $A$ obeys
\begin{equation*}
\gamma_{2k+s-1}<\frac{\sqrt{2}}{4}
\end{equation*}
and
\begin{equation*}
\lambda>\frac{2\sqrt{2}\gamma_{2k+s-1}(1+\delta_{2k+s-1})\sqrt{1+\delta_{2k+s-1}}}{\sqrt{1-\delta_{2k+s-1}}(1-2\sqrt{2}\gamma_{2k+s-1})},
\end{equation*}
then the sequence $\{\mu^{k}\}$ produced by AdptNST+HT+subOptFB with $y=Ax+e$ for $s$-sparse signal $x$ and error $e$ satisfies
\begin{equation*}
\|x-\mu^{k}\|_{2}\leq\rho^{k}_{2k+s-1}\|x-\mu^{0}\|_{2}+\kappa_{2k+s-1}\frac{1-\rho^{k}_{2k+s-1}}{1-\rho_{2k+s-1}}\|e\|_{2},
\end{equation*}
where $\rho_{s}=2\sqrt{2}\gamma_{s}(1+\frac{(1+\delta_{s})\sqrt{1+\delta_{s}}}{\lambda\sqrt{1-\delta_{s}}})$ and $\kappa_{s}=(1+\frac{(1+\delta_{s})\sqrt{1+\delta_{s}}}{\lambda\sqrt{1-\delta_{s}}})(\sqrt{2+2\theta_{s}}+\sqrt{1+\theta_{s}})+\frac{(1+\delta_{s})}{\lambda\sqrt{1-\delta_{s}}}$.
\end{theorem}

$\bm{Proof.}$
 Let supp$(x)=S$, $R_{k}=S\cup T_{k}$, where $|T_{k}|=k$ and $v^{k}_{T_{k}}=x^{k}_{T_{k}}$, $v^{k}_{T_{k}^{c}}=0$.
Since $\mu_{T_{k}}^{k}=x_{T_{k}}^{k}+\lambda^{-1}A_{T_{k}}^{\ast}A_{T_{k}^{c}}x_{T_{k}^{c}}^{k}$, and the feasibility of $x^{k}$ (i.e., $y=A_{T_{k}}x_{T_{k}}^{k}+A_{T_{k}^{c}}x_{T_{k}^{c}}^{k}$),
\begin{equation*}
\lambda(\mu_{T_{k}}^{k}-x_{T_{k}}^{k})=A_{T_{k}}^{\ast}(Ax-Av^{k})+A_{T_{k}}^{\ast}e.\\
\end{equation*}
Consequently,
\begin{equation*}
\lambda A_{T_{k}}(\mu_{T_{k}}^{k}-x_{T_{k}}^{k})=A_{T_{k}}A_{T_{k}}^{\ast}(Ax-Av^{k})+A_{T_{k}}A_{T_{k}}^{\ast}e.\\
\end{equation*}
It implies that
\begin{equation*}
\lambda\|A(\mu^{k}-v^{k})\|_{2}\leq\|A_{T_{k}}A_{T_{k}}^{\ast}A(x-v^{k})\|_{2}+\|A_{T_{k}}A_{T_{k}}^{\ast}e\|_{2}.
\end{equation*}
The left-hand satisfies
\begin{equation*}
\begin{aligned}
\lambda\|A(\mu^{k}-v^{k})\|_{2}\geq\lambda\sqrt{1-\delta_{k}}\|\mu^{k}-v^{k}\|_{2},
\end{aligned}
\end{equation*}
while the right-hand side satisfies
\begin{equation*}
\begin{aligned}
&\|A_{T_{k}}A_{T_{k}}^{\ast}A(x-v^{k})\|_{2}+\|A_{T_{k}}A_{T_{k}}^{\ast}e\|_{2}\\
&\leq\|A_{T_{k}}A_{T_{k}}^{\ast}\|_{2}\|A(x-v^{k})\|_{2}+\|A_{T_{k}}A_{T_{k}}^{\ast}\|_{2}\|e\|_{2}\\
&\leq(1+\delta_{k})\sqrt{1+\delta_{s+k}}\|x-v^{k}\|_{2}+(1+\delta_{k})\|e\|_{2}.\\
\end{aligned}
\end{equation*}
The last step is due to $1-\delta_{k}\leq\|A_{T_{k}}A_{T_{k}}^{\ast}\|_{2}=\|A_{T_{k}}^{\ast}A_{T_{k}}\|_{2}\leq1+\delta_{k}$.

\noindent Therefore, one has
\begin{equation*}
\begin{aligned}
&\|\mu^{k}-v^{k}\|_{2}\\
&\leq\frac{(1+\delta_{k})\sqrt{1+\delta_{s+k}}}{\lambda\sqrt{1-\delta_{k}}}\|x-v^{k}\|_{2}+\frac{(1+\delta_{k})}{\lambda\sqrt{1-\delta_{k}}}\|e\|_{2}.\\
\end{aligned}
\end{equation*}
Using the triangle inequality, we further derive
\begin{equation}\label{eq18}
\begin{aligned}
&\|\mu^{k}-x\|_{2}\\
&\leq\|\mu^{k}-v^{k}\|_{2}+\|v^{k}-x\|_{2}\\
&\leq(1+\frac{(1+\delta_{k})\sqrt{1+\delta_{s+k}}}{\lambda\sqrt{1-\delta_{k}}})\|x-v^{k}\|_{2}+\frac{(1+\delta_{k})}{\lambda\sqrt{1-\delta_{k}}}\|e\|_{2}.\\
\end{aligned}
\end{equation}

It then follows that
\begin{equation*}
\begin{aligned}
&\|x-v^{k}\|_{2}\\
&\leq\|(x-v^{k})_{T_{k}}\|_{2}+\|(x-v^{k})_{R_{k}\backslash T_{k}}\|_{2}\\
&=\|x_{T_{k}}-x^{k}_{T_{k}}\|_{2}+\|x_{S\backslash T_{k}}\|_{2}\\
&=\|[\mu^{k-1}+A^{\ast}(AA^{\ast})^{-1}(y-A\mu^{k-1})]_{T_{k}}-x_{T_{k}}\|_{2}+\|x_{S\backslash T_{k}}\|_{2}\\
&=\|\mu^{k-1}_{T_{k}}-x_{T_{k}}+[A^{\ast}(AA^{\ast})^{-1}(Ax+e-A\mu^{k-1})]_{T_{k}}\|_{2}\\
&+\|x_{S\backslash T_{k}}\|_{2}\\
&\leq\|\mu^{k-1}_{T_{k}}-x_{T_{k}}+A^{\ast}_{T_{k}}(AA^{\ast})^{-1}(Ax-A\mu^{k-1})\|_{2}\\
&+\|[A^{\ast}(AA^{\ast})^{-1}e]_{T_{k}}\|_{2}+\|x_{S\backslash T_{k}}\|_{2}\\
&=\|\mu^{k-1}_{T_{k}}-x_{T_{k}}+A^{\ast}_{T_{k}}(AA^{\ast})^{-1}A_{T_{k}}(x-\mu^{k-1})_{T_{k}} \\
& + A^{\ast}_{T_{k}}(AA^{\ast})^{-1}A_{R_{k-1}\backslash T_{k}}(x-\mu^{k-1})_{R_{k-1}\backslash T_{k}}\|_{2}\\ &+\|[A^{\ast}(AA^{\ast})^{-1}e]_{T_{k}}\|_{2}+\|x_{S\backslash T_{k}}\|_{2}\\
&\leq\|\mu^{k-1}_{T_{k}}-x_{T_{k}}\\
&+A^{\ast}_{T_{k}}(AA^{\ast})^{-1}A_{T_{k}}(x-\mu^{k-1})_{T_{k}}\|_{2}\\
&+\|A^{\ast}_{T_{k}}(AA^{\ast})^{-1}A_{R_{k-1}\backslash T_{k}}(x-\mu^{k-1})_{R_{k-1}\backslash T_{k}}\|_{2}\\
&+\|[A^{\ast}(AA^{\ast})^{-1}e]_{T_{k}}\|_{2}+\|x_{S\backslash T_{k}}\|_{2}\\
&=\|(I-A^{\ast}_{T_{k}}(AA^{\ast})^{-1}A_{T_{k}})(\mu^{k-1}-x)_{T_{k}}\|_{2}\\
&+\|A^{\ast}_{T_{k}}(AA^{\ast})^{-1}A_{R_{k-1}\backslash T_{k}}(x-\mu^{k-1})_{R_{k-1}\backslash T_{k}}\|_{2}\\
&+\|[A^{\ast}(AA^{\ast})^{-1}e]_{T_{k}}\|_{2}+\|x_{S\backslash T_{k}}\|_{2}\\
&\leq\gamma_{k}\|(\mu^{k-1}-x)_{T_{k}}\|_{2}+\gamma_{2k+s-1}\|(x-\mu^{k-1})_{R_{k-1}\backslash T_{k}}\|_{2}\\
&+\sqrt{1+\theta_{k}}\|e\|_{2}\\
&+\sqrt{2}(\gamma_{2k+s-1}\|x-\mu^{k-1}\|_{2}+\sqrt{1+\theta_{k+s}}\|e\|_{2}),
\end{aligned}
\end{equation*}
where the last step is due to {\em Definition \ref{de2}}, {\em Lemma \ref{le2}}, {\em Lemma \ref{le3}} and {\em Lemma \ref{le4}}. $(\mu^{k-1}-x)_{T_{k}}$ and $(x-\mu^{k-1})_{R_{k-1}\backslash T_{k}}$ are orthogonal so that $\|(\mu^{k-1}-x)_{T_{k}}\|_{2}+\|(x-\mu^{k-1})_{R_{k-1}\backslash T_{k}}\|_{2}\leq\sqrt{2}\|x-\mu^{k-1}\|_{2}$. Since $\gamma_{k}\leq\gamma_{2k+s-1}$, we therefore have
\begin{equation}\label{eq19}
\begin{aligned}
&\|x-v^{k}\|_{2}\\
&\leq2\sqrt{2}\gamma_{2k+s-1}\|x-\mu^{k-1}\|_{2}+(\sqrt{2+2\theta_{k+s}}+\sqrt{1+\theta_{k}})\|e\|_{2}.\\
\end{aligned}
\end{equation}
Combining (\ref{eq18}) and (\ref{eq19}), it yields
\begin{equation*}
\begin{aligned}
&\|x-\mu^{k}\|_{2}\\
&\leq2\sqrt{2}\gamma_{2k+s-1}(1+\frac{(1+\delta_{k})\sqrt{1+\delta_{s+k}}}{\lambda\sqrt{1-\delta_{k}}})\|x-\mu^{k-1}\|_{2}\\
&+[(1+\frac{(1+\delta_{k})\sqrt{1+\delta_{s+k}}}{\lambda\sqrt{1-\delta_{k}}})(\sqrt{2+2\theta_{k+s}}+\sqrt{1+\theta_{k}})\\
&+\frac{(1+\delta_{k})}{\lambda\sqrt{1-\delta_{k}}}]\|e\|_{2}\\
&\leq\rho_{2k+s-1}\|x-\mu^{k-1}\|_{2}+\kappa_{s+k}\|e\|_{2}.
\end{aligned}
\end{equation*}
The sequence $\{\mu^{k}\}$ converges if $2\sqrt{2}\gamma_{2k+s-1}(1+\frac{(1+\delta_{k})\sqrt{1+\delta_{s+k}}}{\lambda\sqrt{1-\delta_{k}}})<1$, i.e.,
\begin{equation*}
\gamma_{2k+s-1}<\frac{\sqrt{2}}{4}
\end{equation*}
and
\begin{equation*}
\lambda>\frac{2\sqrt{2}\gamma_{2k+s-1}(1+\delta_{2k+s-1})\sqrt{1+\delta_{2k+s-1}}}{\sqrt{1-\delta_{2k+s-1}}(1-2\sqrt{2}\gamma_{2k+s-1})},
\end{equation*}
which provides the asserted convergence condition.
In turn, we can derive that
\begin{equation*}
\|x-\mu^{k}\|_{2}\leq\rho^{k}_{2k+s-1}\|x-\mu^{0}\|_{2}+\kappa_{2k+s-1}\frac{1-\rho^{k}_{2k+s-1}}{1-\rho_{2k+s-1}}\|e\|_{2}.
\end{equation*}

\begin{corollary}
In particular, if the sparsity $s$ is known $(|T_{k}|=s)$, it is easy to get the convergence result of NST+HT+subOptFB.
\noindent Let $x$ be the solution to $y=Ax+e$ with sparsity $s$. If the P-RIP of $A$ satisfies
\begin{equation*}
\gamma_{3s}<\frac{\sqrt{2}}{4},
\end{equation*}
and
\begin{equation*}
 \lambda>\frac{2\sqrt{2}\gamma_{3s}(1+\delta_{3s})\sqrt{1+\delta_{3s}}}{\sqrt{1-\delta_{3s}}(1-2\sqrt{2}\gamma_{3s})} ,
 \end{equation*}
 then the sequence $\{\mu^{k}\}$ $(k\geq 1)$ produced by NST+HT+subOptFB satisfies
\begin{equation*}
\|x-\mu^{k}\|_{2}\leq\rho^{k}_{3s}\|x-\mu^{0}\|_{2}+\kappa_{3s}\frac{1-\rho^{k}_{3s}}{1-\rho_{3s}}\|e\|_{2}.
\end{equation*}
Furthermore, for Parseval frames, since $AA^{\ast}=I$, $\gamma_{s}(A)=\delta_{s}(A)$, and $\theta_{s}(A)=\delta_{s}(A)$, one can obtain the following convergence result.

\noindent Let $x$ be the solution to $y=Ax+e$ with $s$ sparsity. If the RIP of the Parseval frame $A$ satisfies
\begin{equation*}
 \delta_{3s}<\frac{\sqrt{2}}{4},
\end{equation*}
 and
\begin{equation*}
 \lambda>\frac{2\sqrt{2}\delta_{3s}(1+\delta_{3s})\sqrt{1+\delta_{3s}}}{\sqrt{1-\delta_{3s}}(1-2\sqrt{2}\delta_{3s})} ,
 \end{equation*}
 then the sequence $\{\mu^{k}\}$ $(k\geq 1)$ produced by NST+HT+subOptFB satisfies
\begin{equation*}
\|x-\mu^{k}\|_{2}\leq\rho^{k}_{3s}\|x-\mu^{0}\|_{2}+\kappa_{3s}\frac{1-\rho^{k}_{3s}}{1-\rho_{3s}}\|e\|_{2}.
\end{equation*}
\noindent In this case, for $\delta_{3s}\leq\frac{1}{4}$, $\lambda>3.9$ can guarantee the convergence.
\end{corollary}
\subsection{Computational Complexity}
It is instructive to examine the computational complexity of different approaches at each iteration. For the null space tuning step, $A^{\ast}(AA^{\ast})^{-1}$ does not change the appearance during iterations.  Consequently, if the inversion $(AA^{\ast})^{-1}$ is calculated off-line, then $A^{\ast}(AA^{\ast})^{-1}$ can be stored in the memory. The computational complexity of the null space tuning is $\mathcal {O}(NM)$ per iteration.

For NST+HT+FB, the update of $\mu^{k}$ takes $\mathcal {O}(s^{3}+MN)$. Without the knowledge of sparsity, the procedure takes $\mathcal {O}(k^{3}+MN)$, where the sparsity $k$ increases gradually. The suboptimal feedbacks avoids calculating the matrix inversion, the computational complexity of updating $\mu^{k}$ can reduce to $\mathcal {O}(s+MN)$. For AdptNST+HT+subOptFB, the update of $\mu^{k}$ takes $\mathcal {O}(k+MN)$, where $k$ is the sparsity at each iteration. For another state-of-the-art iterative thresholding algorithm called the hard thresholding pursuit algorithm (HTP) \cite{foucart2011hard}, the computational complexity of choosing the support is $\mathcal {O}(MN)$, while the update of $x^{k}$ takes $\mathcal {O}(s^{3}+sM)$. The computational complexity of updating $x^{k}$ in the graded hard thresholding pursuit algorithm (GHTP) \cite{bouchot2016hard} is therefore $\mathcal {O}(k^{3}+kM)$, where $|T_{k}|=k$ at each iteration. As a result, AdptNST+HT+subOptFB is perhaps the most efficient one among afore-mentioned  algorithms.
\begin{figure}[t]
  \centering
  \subfigure[]{
  \begin{minipage}[b]{0.30\textwidth}
  \label{fig: parameters-a}\includegraphics[width=1.1\textwidth]{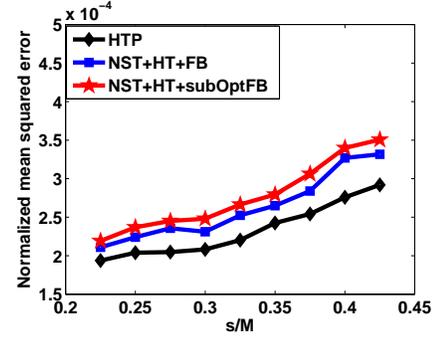}
  \end{minipage}}
  \subfigure[]{
  \begin{minipage}[b]{0.30\textwidth}
  \label{fig: parameters-a}\includegraphics[width=1.1\textwidth]{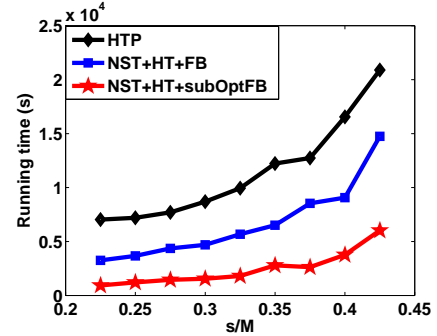}
  \end{minipage}}
  \caption{NMSE(a) The running time(b) versus  the sparsity ratio $s/M$. Here the signal dimension $N=100000$ and the undersampling ratio $M/N=0.5$. All algorithms recover the sparse signal with the prior sparsity value $s$.}
  \end{figure}
\begin{figure}[t]
  \centering
  \subfigure[]{
  \begin{minipage}[b]{0.30\textwidth}
  \label{fig: parameters-a}\includegraphics[width=1.1\textwidth]{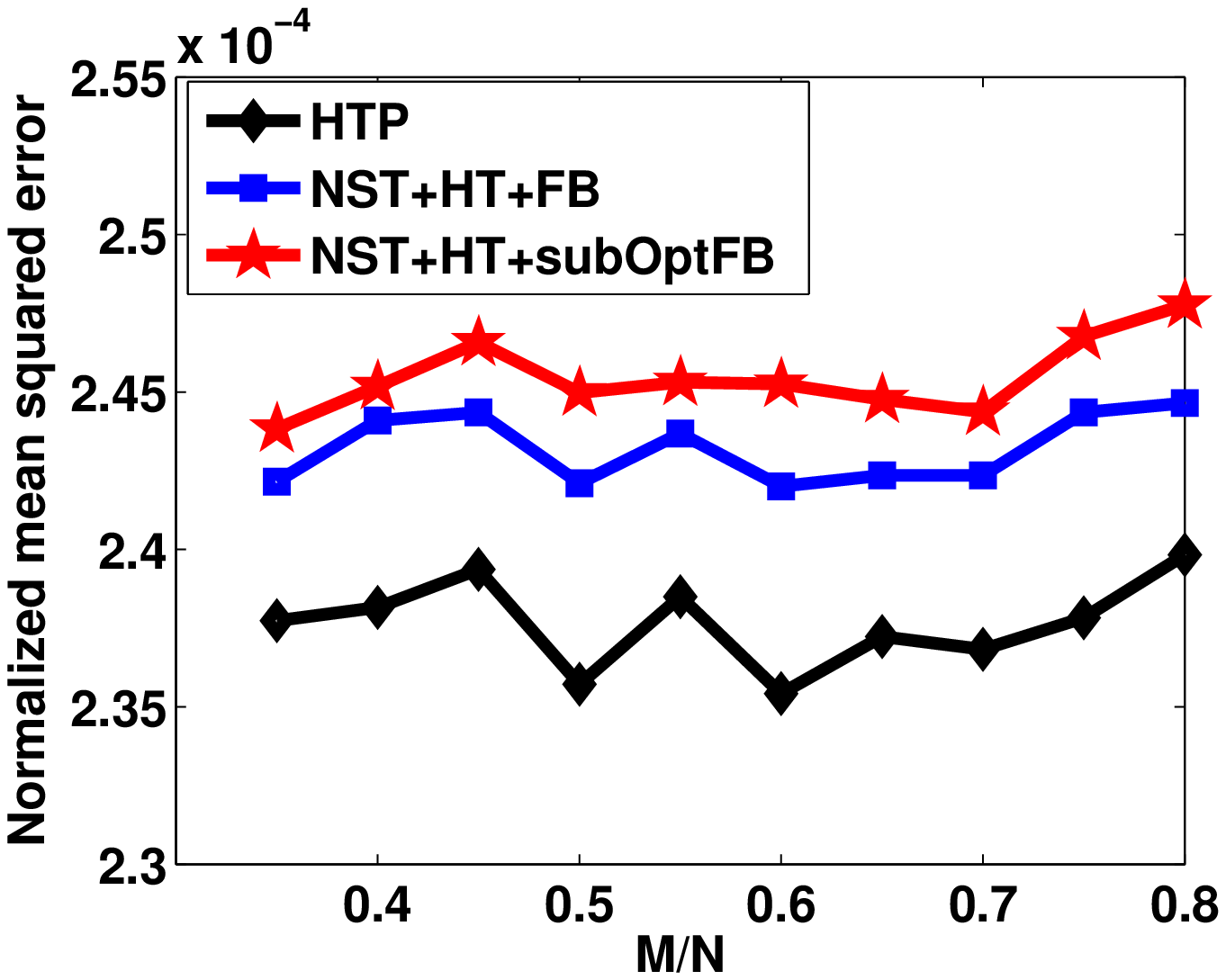}
  \end{minipage}}
  \subfigure[]{
  \begin{minipage}[b]{0.30\textwidth}
  \label{fig: parameters-a}\includegraphics[width=1.1\textwidth]{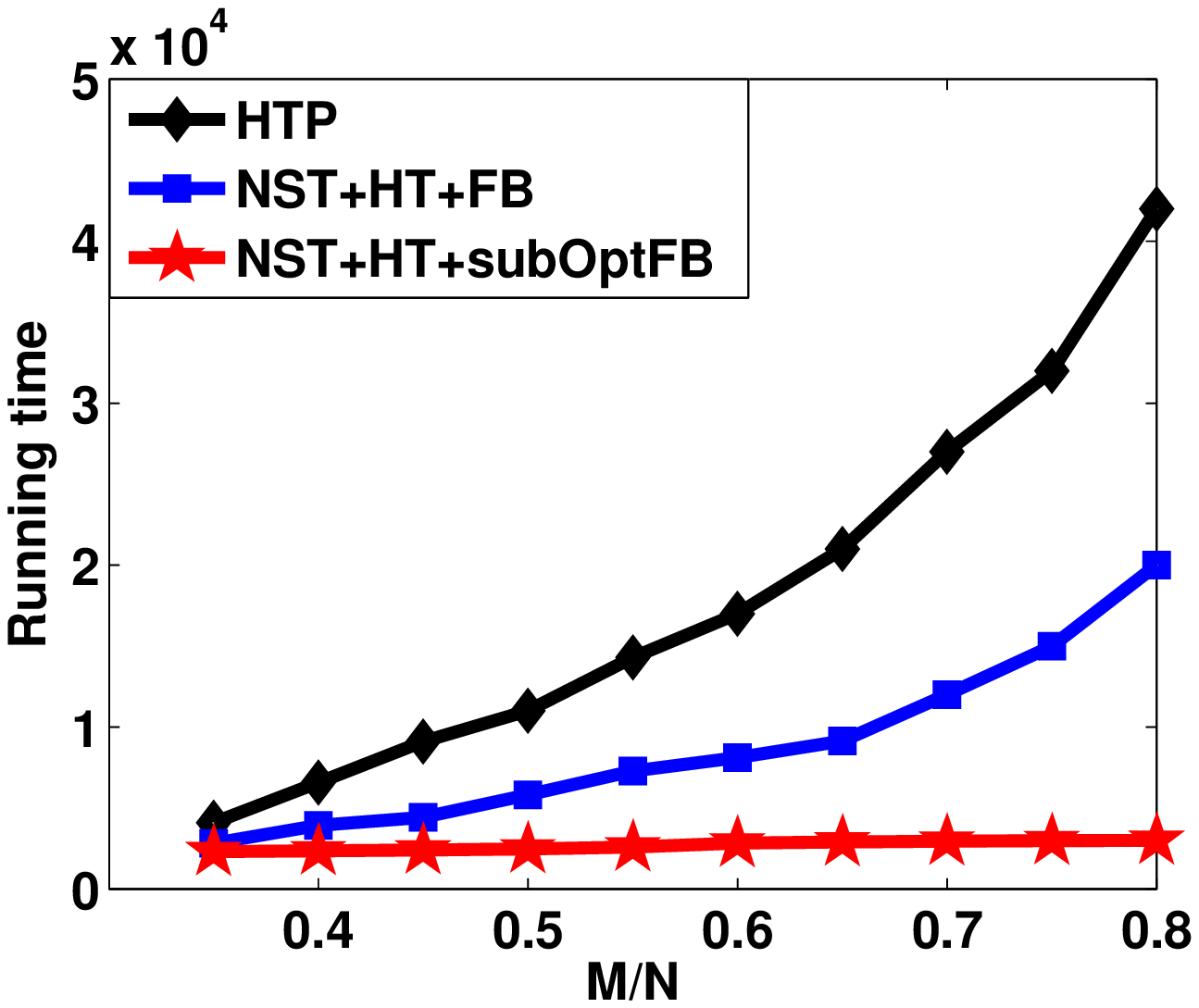}
  \end{minipage}}
  \caption{NMSE(a) The running time(b) versus the undersampling ratio $M/N$. Here the signal dimension $N=100000$ and the sparsity ratio $s/M=0.3$. All algorithms recover the sparse signal with the prior sparsity value $s$.}
  \end{figure}
\section{Numerical experiments}
This section provides several experiments to evaluate the performances of NST+HT+subOptFB and AdptNST+HT+subOptFB, including (1) simulations with the knowledge of sparsity level (2) their adaptive versions without the prior estimation of sparsity level. We focus on comparing the performance of NST+HT+subOptFB with two state-of-the-art iterative thresholding algorithms, i.e., HTP, NST+HT+FB, in terms of large scale problems. All experimental results are executed on a $3.50$ GHZ Intel core CPU and $500$GB memory.

Two metrics are considered to evaluate the recovery performance of respective algorithms. The first metric is the normalized mean squared errors (NMSE), which is calculated by averaging normalized squared errors $\|x-\widehat{x}\|_{2}^{2}/\|x\|_{2}^{2}$ over independent trials. $\widehat{x}$ denotes the estimation of the true signal $x$. The second metric gauge running times, which is the most important metric given the prevalence of large scale datasets. Considering the real case, we add small white Gaussian noise to the real data. The compared experiments with the prior estimation of sparsity value refer to HTP, NST+HT+FB, and NST+HT+subOptFB, while GHTP, AdptNST+HT+FB and  AdptNST+HT+subOptFB are related to signals without the knowledge of sparsity value.
\begin{figure}[t]
  \centering
  \subfigure[]{
  \begin{minipage}[b]{0.30\textwidth}
  \label{fig: parameters-a}\includegraphics[width=1.1\textwidth]{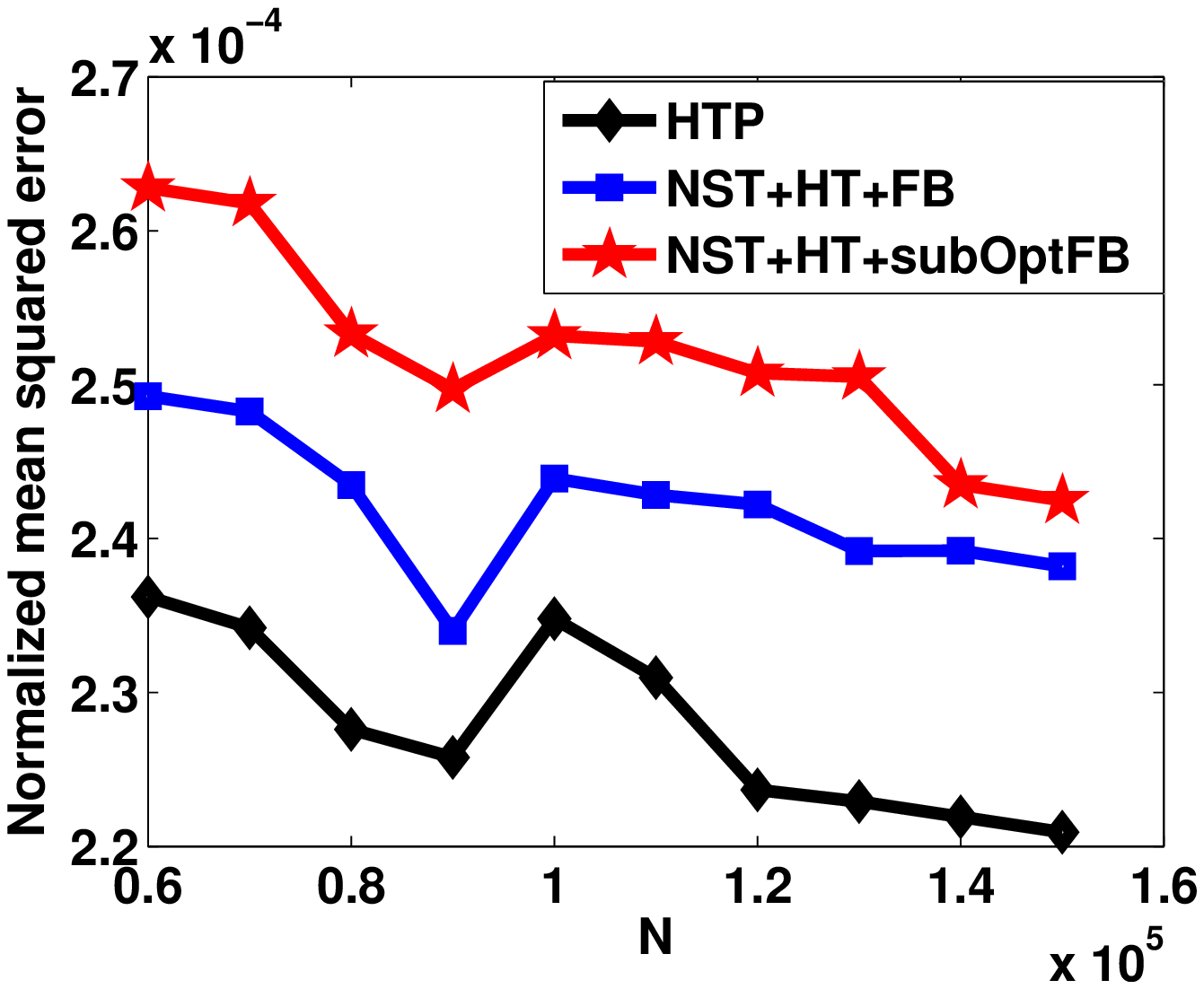}
  \end{minipage}}
  \subfigure[]{
  \begin{minipage}[b]{0.30\textwidth}
  \label{fig: parameters-a}\includegraphics[width=1.1\textwidth]{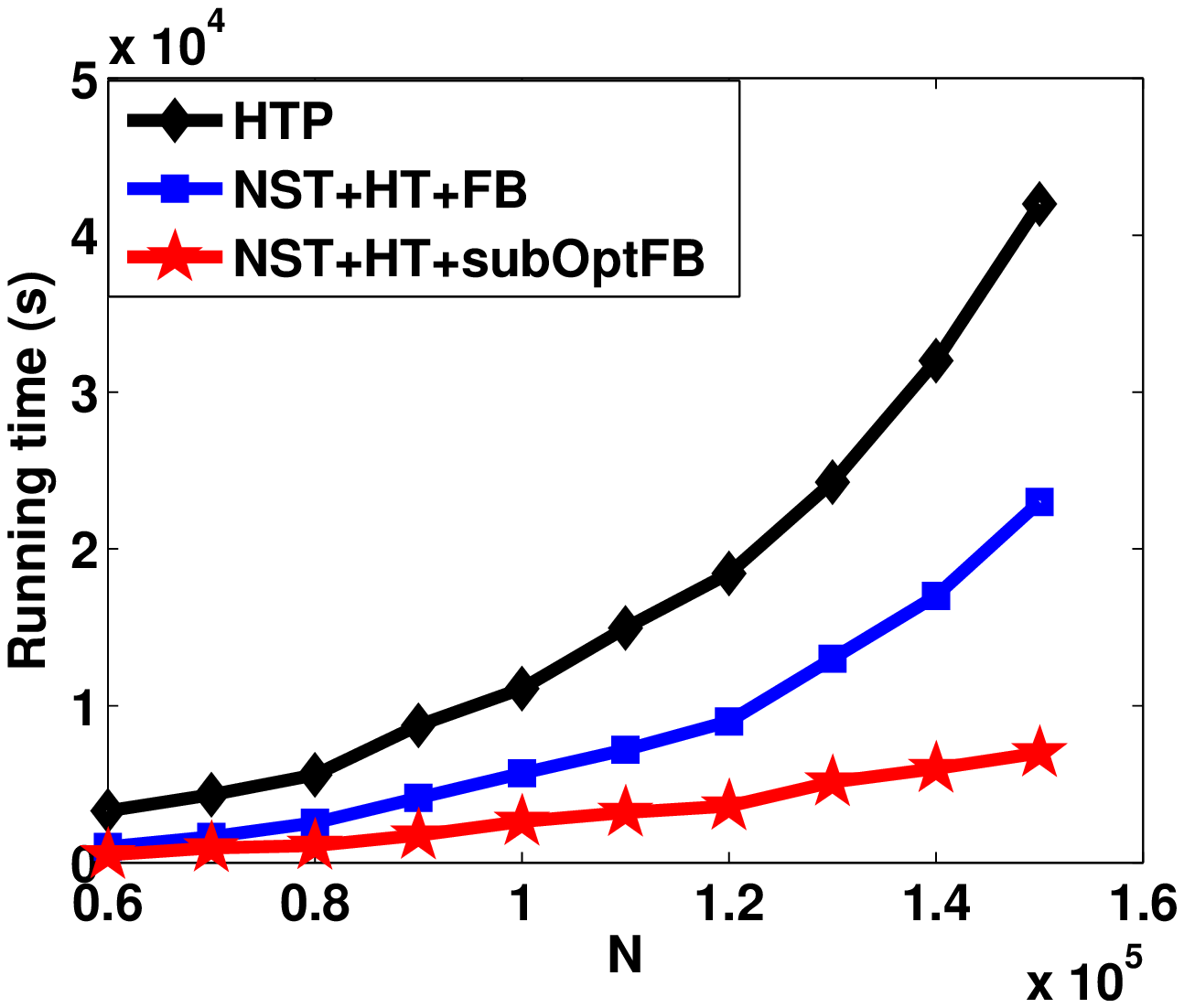}
  \end{minipage}}
  \caption{NMSE(a) The running time(b) versus the signal dimension $N$. Here the undersampling ratio $M/N=0.5$ and the sparsity ratio $s/M=0.3$. All algorithms recover the sparse signal with the prior sparsity value $s$.}
  \end{figure}
\begin{figure}[t]
  \centering
  \subfigure[]{
  \begin{minipage}[b]{0.30\textwidth}
  \label{fig: parameters-a}\includegraphics[width=1.1\textwidth]{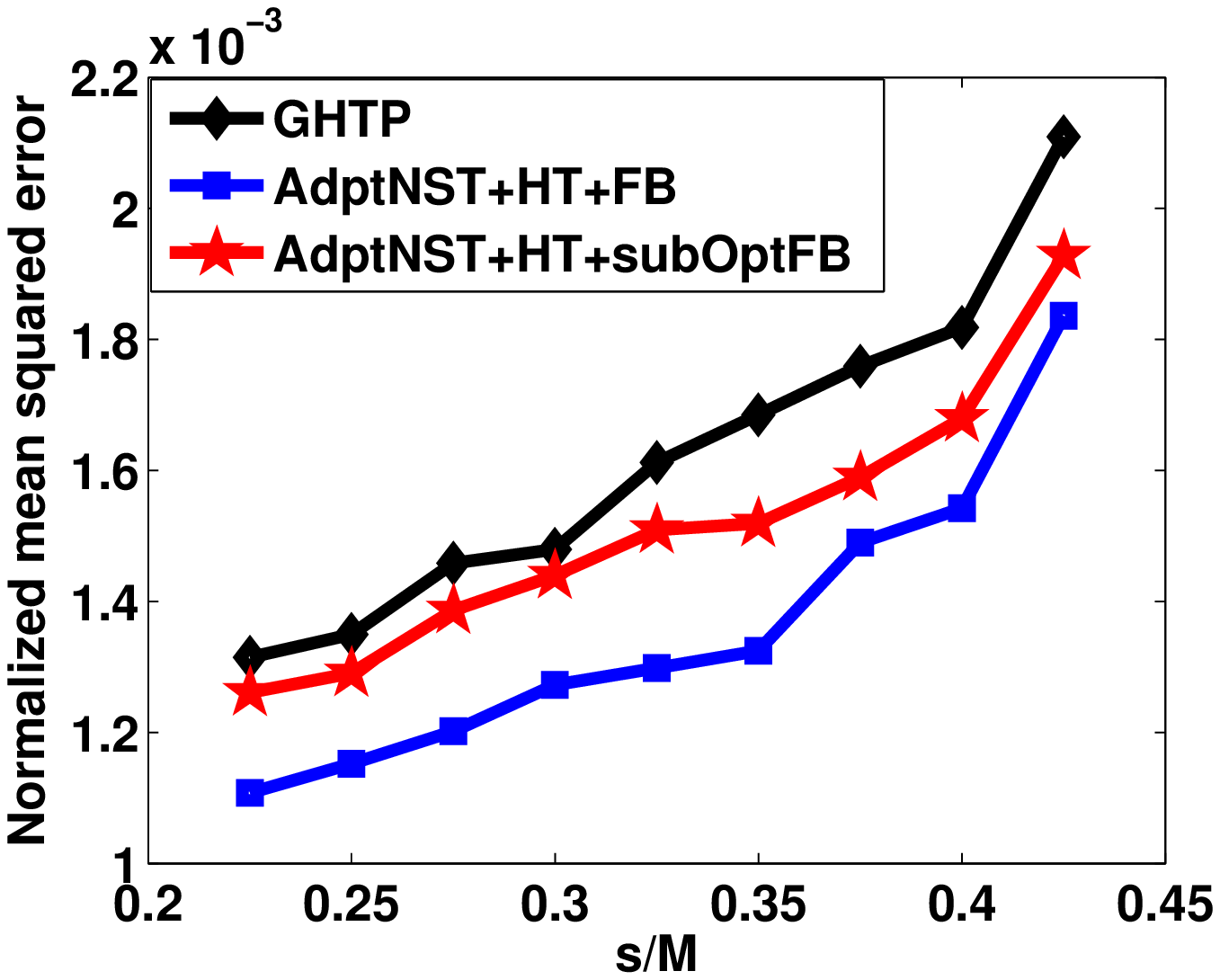}
  \end{minipage}}
  \subfigure[]{
  \begin{minipage}[b]{0.30\textwidth}
  \label{fig: parameters-a}\includegraphics[width=1.1\textwidth]{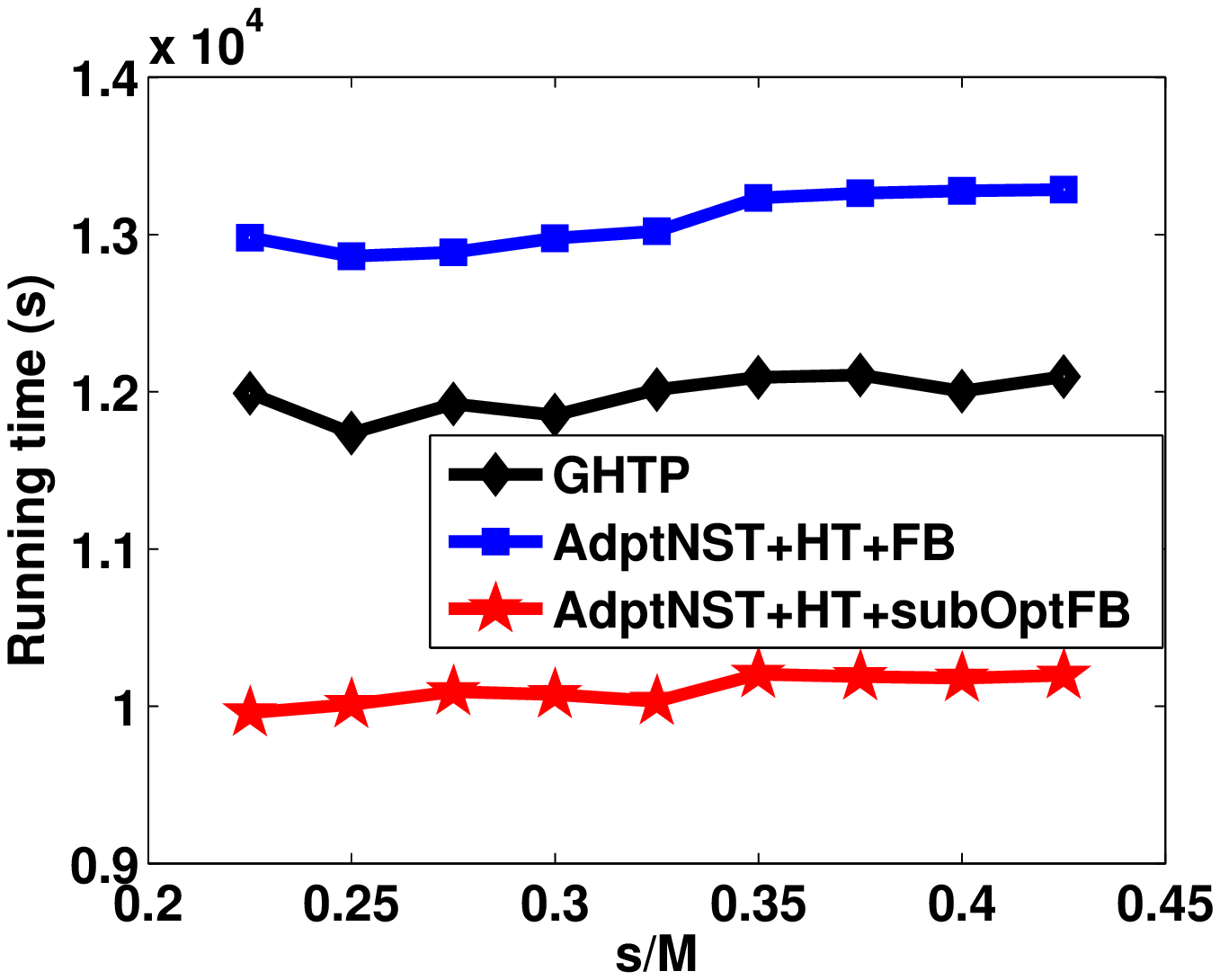}
  \end{minipage}}
  \caption{NMSE(a) The running time(b) versus the sparsity ratio $s/M$. Here the signal dimension $N=10000$ and the undersampling ratio $M/N=0.5$. All algorithms recover the sparse signal without the prior sparsity value $s$.}
 \end{figure}

\begin{figure}[t]
  \centering
  \subfigure[]{
  \begin{minipage}[b]{0.30\textwidth}
  \label{fig: parameters-a}\includegraphics[width=1.1\textwidth]{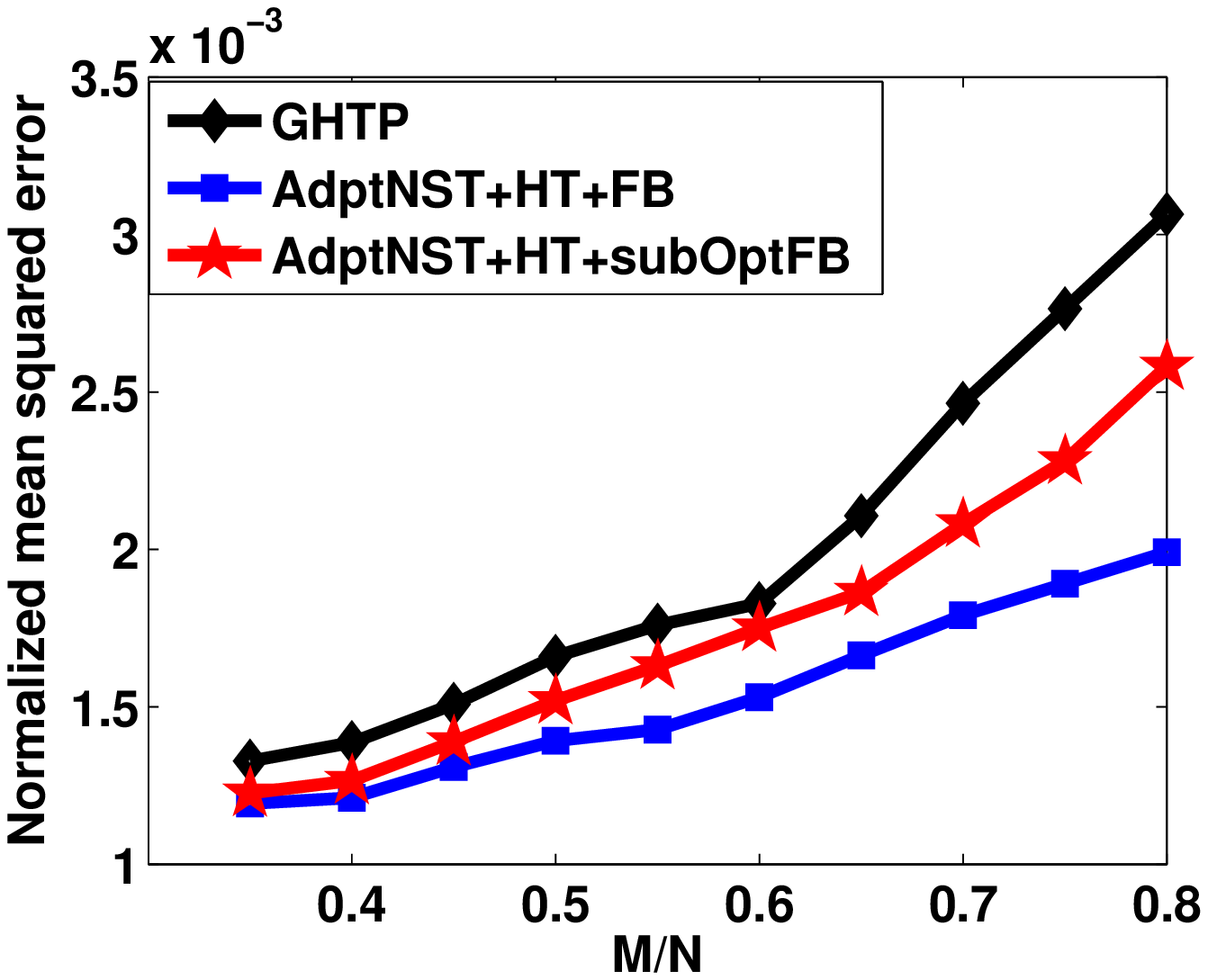}
  \end{minipage}}
  \subfigure[]{
  \begin{minipage}[b]{0.30\textwidth}
  \label{fig: parameters-a}\includegraphics[width=1.1\textwidth]{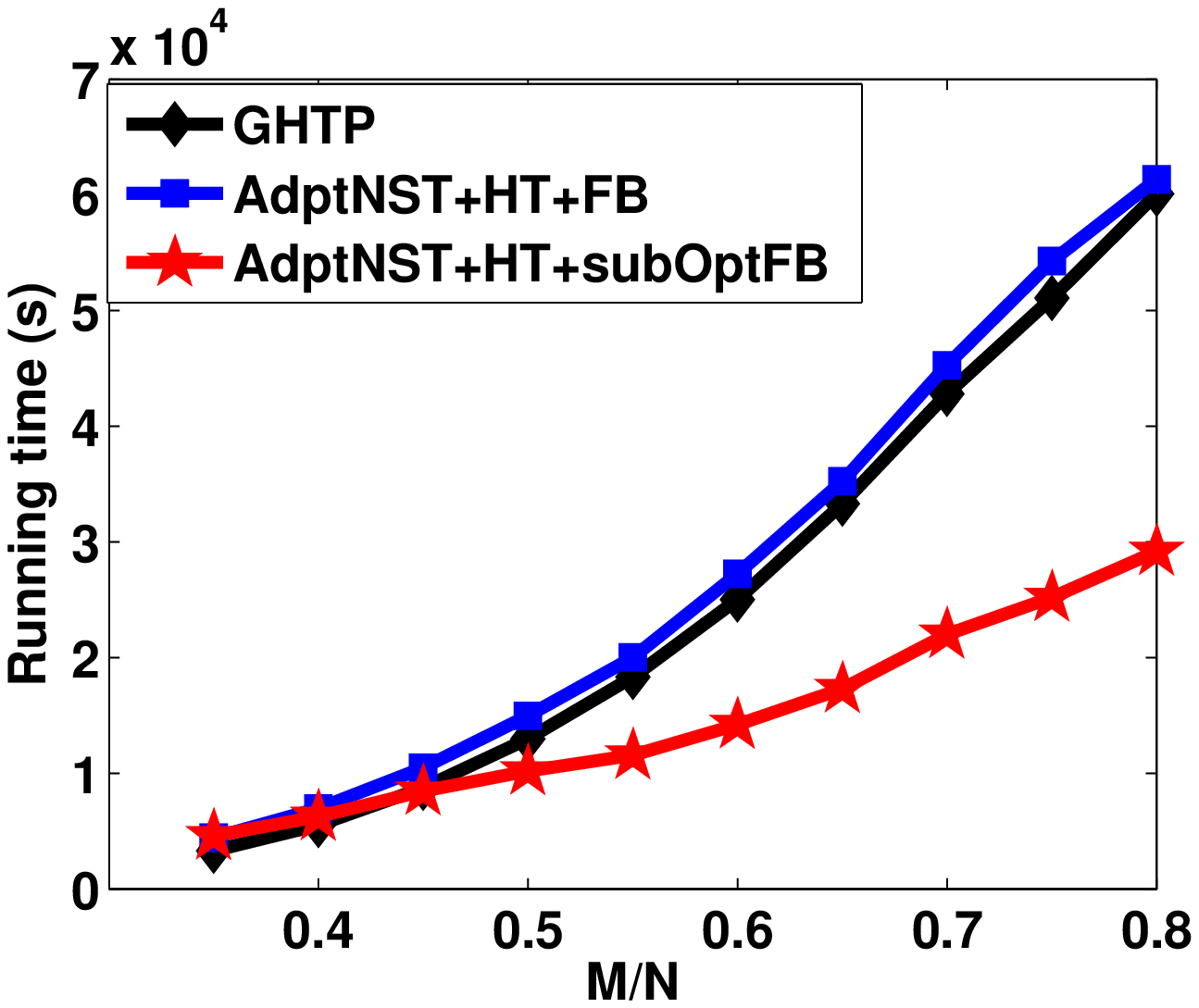}
  \end{minipage}}
  \caption{NMSE(a) The running time(b) versus the undersampling ratio $M/N$. Here the signal dimension $N=10000$ and the sparsity ratio $s/M=0.3$. All algorithms recover the sparse signal without the prior sparsity value $s$.}
  \end{figure}
\subsection{Signals with the knowledge of sparsity level}
\begin{figure}[t]
  \centering
  \subfigure[]{
  \begin{minipage}[b]{0.30\textwidth}
  \label{fig: parameters-a}\includegraphics[width=1.1\textwidth]{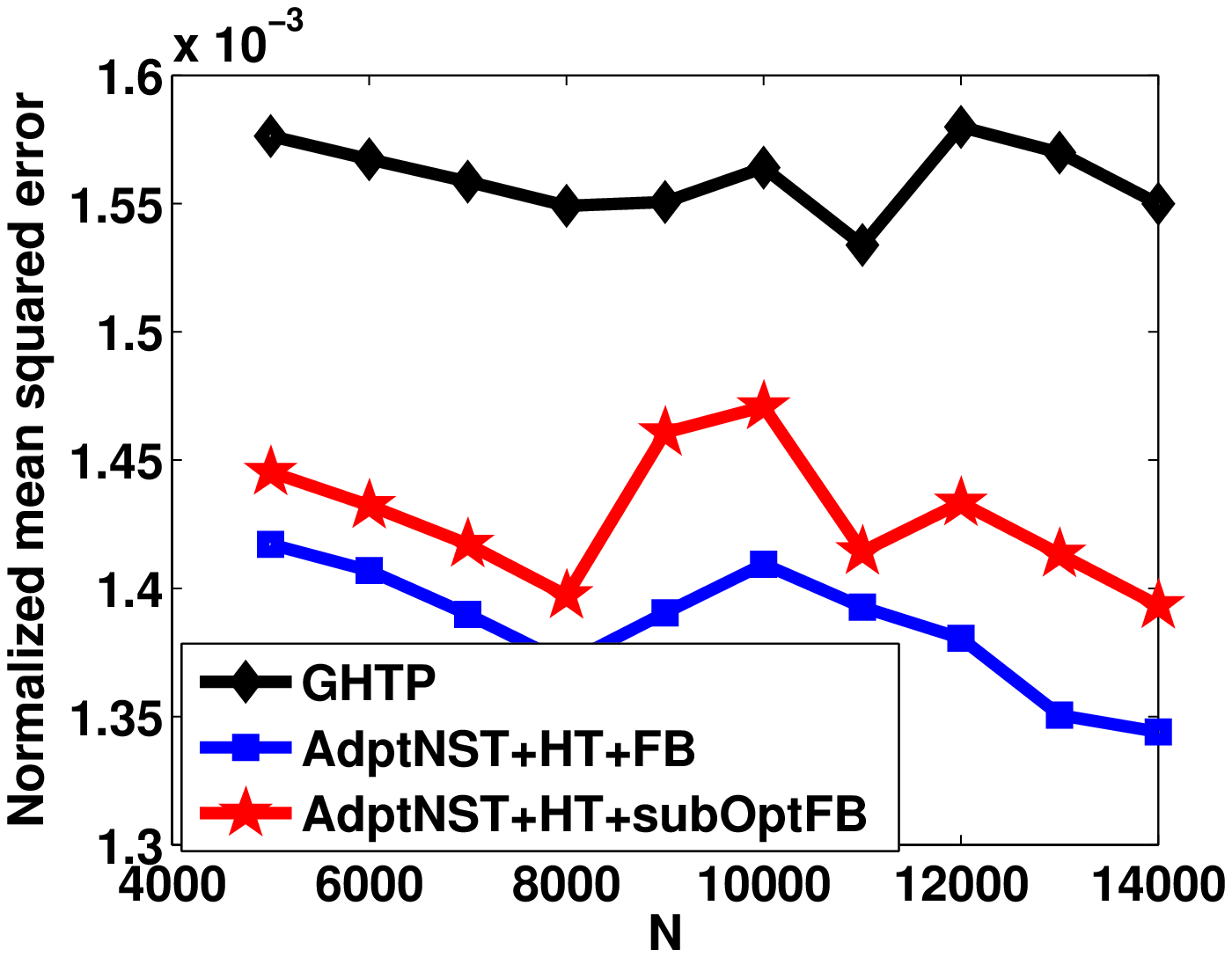}
  \end{minipage}}
  \subfigure[]{
  \begin{minipage}[b]{0.30\textwidth}
  \label{fig: parameters-a}\includegraphics[width=1.1\textwidth]{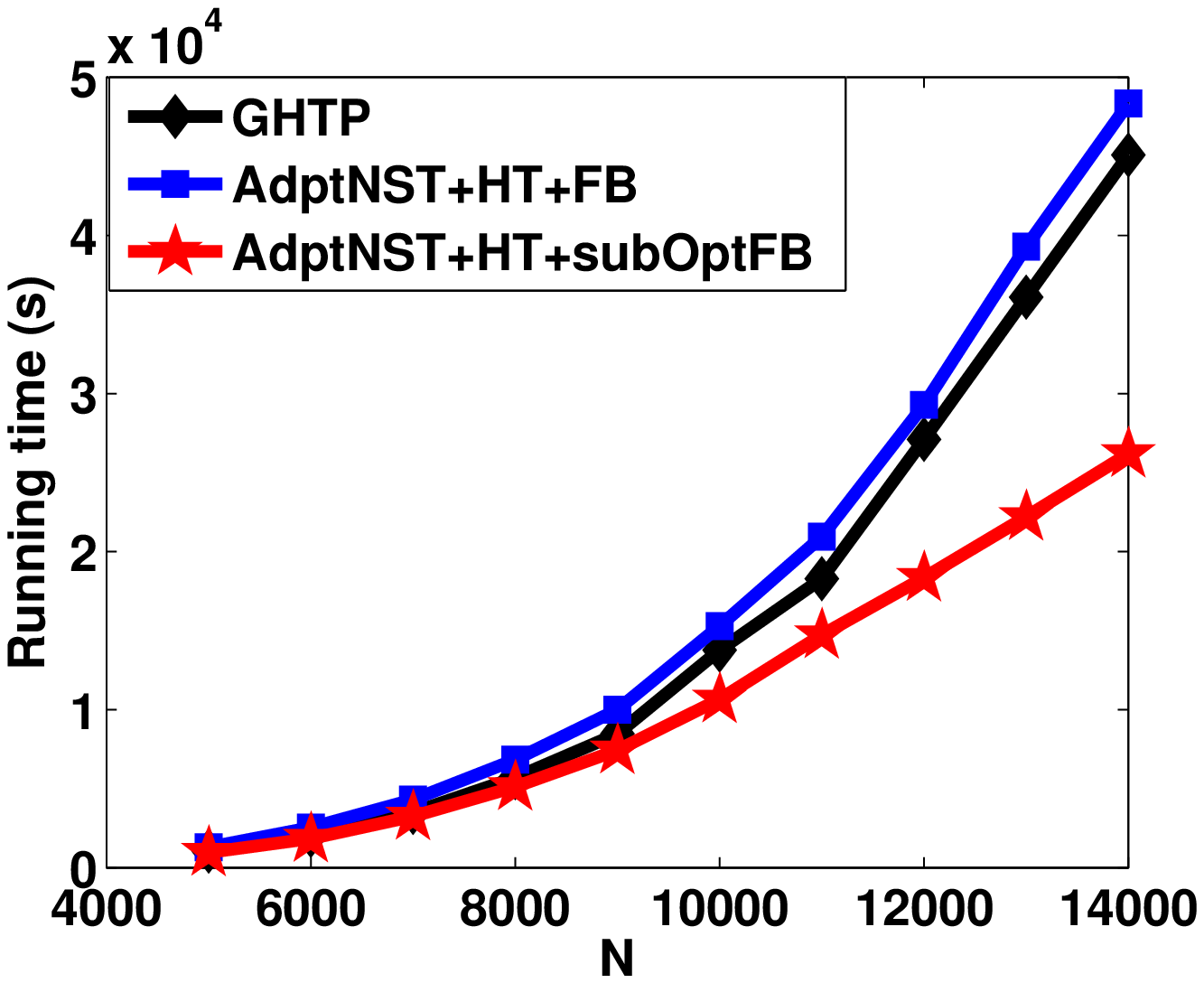}
  \end{minipage}}
  \caption{NMSE(a) The running time(b) versus the signal dimension $N$. Here the undersampling ratio $M/N=0.5$ and the sparsity ratio $s/M=0.3$. All algorithms recover the sparse signal without the prior sparsity value $s$.}
  \end{figure}
As a first experiment, we study how performance changes as a function of the measurements-to-active-coefficients ratio, $s/M$. In this experiment, $N=100000$, $M=50000$. The ratio of measurements-to-active-coefficients, $s/M$ ranges from $0.225$ to $0.425$. {\em Fig.1 (a)} shows that all algorithms degrade with increasing $s/M$. There is no notable difference in NMSE among different algorithms as $s/M$ increases. As indicated in {\em Fig.1 (b)},
NST+HT+subOptFB is computationally considerably more efficient compared to other methods.

 We then study how performance changes as a function of the undersampling ratio, $M/N$. In {\em Fig.2}, we present the results in which the undersampling ratio, $M/N$, is varied from $0.35$ to $0.8$. Specially, $N$ is fixed at $100000$, while $M$ is varied. The sparsity ratio of active-coefficients-measurements $s/M$ is fixed at $0.3$. Since the sparsity $s$ increases with increasing $M/N$, there is no notable characters of NMSE with varying $M/N$. It is clear that all methods provide very good reconstructions with high reconstructed accuracy,  ranged from $2.35\times10^{-4}$ to $2.48\times10^{-4}$. Referring to running times, the slope of the
running time for NST+HT+subOptFB is much smaller than those of other methods. It is concluded from the figure that NST+HT+subOptFB is the most efficient algorithm, which is much faster without compromising the recovery accuracy.

As discussed throughout this paper, a key consideration of NST+HT+subOptFB is ensuring that it would be suitable for large scale problems. To demonstrate the case, we conduct an experiment by varying the signal scale $N$. $M$ is scaled proportionally so that $M/N=0.5$ and $s/M$ is set to $0.3$. {\em Fig.3 (a)} shows that HTP delivers better NMSE than other methods. As shown in {\em Fig.3 (b)}, the computational costs of all methods increase as the dimension of the signals becomes larger. When $N\leq100000$, HTP and NST+HT+FB remain close to NST+HT+subOptFB. However, HTP and NST+HT+FB quickly depart from NST+HT+subOptFB when $N>100000$. It reflects the superiority of NST+HT+subOptFB in dealing with large scale problems.

In summary, we can conclude that HTP delivers good performance in NMSE, which is slightly better than ones of NST+HT+FB and NST+HT+subOptFB. NST+HT+subOptFB provides superior efficiency against other methods at the same level of recovery accuracy. As a result, NST+HT+subOptFB scales with increasing problem dimensions more favorably than HTP and NST+HT+FB. This is not surprising since NST+HT+subOptFB has perfect performance in computational complexity without the matrix inversion at each iteration.

\subsection{Signals without the knowledge of sparsity level}
All previous algorithms require the knowledge of sparsity level. It seems wishful in most applications. To make all algorithms more applicable, we compare their adaptive versions that the size of the index set increases with the iteration.

Similarly, we first test how performance changes as function of $s/M$. The signal dimension $N$ is set to $10000$, $M/N=0.5$ and $s/M$ varies from $0.225$ to $0.425$. {\em Fig.4} plots the results. It can be seen that the performances of all algorithms degrade with increasing $s/M$. AdptNST+HT+FB provides very low reconstruction error. Compared with GHTP and AdptNST+HT+FB, AdptNST+HT+subOptFB is the most efficient algorithm, which is much faster without compromising the recovery accuracy.

 We then investigate how the performances of different algorithms are affected by the undersampling ratio. For this purpose, the signal scale $N$ is set to $10000$ and $s/M$ is fixed at $0.3$. In conducting this test, one can observe that performances are strongly tied to the sparsity $s$. With fixing $s/M=0.3$, the performances show no improvement with more measurements $M$ since the increase of the sparsity $s$. As shown in {\em Fig.5 (a)}, AdptNST+HT+subOptFB, although not as good AdptNST+HT+FB, still delivers acceptable performance which is better than GHTP. In addition, as indicated in {\em Fig.5 (b)}, we can see the tremendous efficiency of AdptNST+HT+subOptFB. The efficiency is particularly evident as sampling rate increases. The execution-time of AdptNST+HT+subOptFB grows much slower than other algorithms as the sampling rate increases.

The comparisons of numerical results by varying signal scale are shown in {\em Fig.6}. While AdptNST+HT+FB provides very low reconstruction errors, its recovery speed is slower than other methods. When $N>10000$, the running times of GHTP and AdptNST+HT+FB increase sharply,
and the performance of AdptNST+HT+subOptFB shows a modest increase. It demonstrates again the great superiority of suboptimal feedbacks in dealing with large scale problems.

Two observations can be made from the above experiments. First, in terms of NMSE, AdptNST+HT+FB performs well, while GHTP and AdptNST+HT+subOptFB are also at the same level. Second, AdptNST+HT+subOptFB is the most efficient algorithm, which is much faster without compromising the recovery accuracy. The low NMSE and the perfect computational efficiency in dealing with signals without the knowledge of sparsity level demonstrate the excellent superiority of suboptimal feedback in real applications.
\section{Conclusion}
Iterative algorithms based on thresholding, feedback and null space tuning is a powerful tool to reconstruct the sparse signal. In particular, suboptimal feedback accelerates the traditional thresholding methods by avoiding the matrix inversion. This paper considers the convergence guarantees of  NST+HT+subOptFB and its variational version AdptNST+HT+subOptFB (without a prior estimation of the sparsity level). The convergence analysis completely confirms the convergence feature of suboptimal feedback. Experimental results demonstrate AdptNST+HT+subOptFB is exceedingly effective and fast for large scale problems.

\section*{Acknowledgment}

This work was partially supported by the NSF of USA (DMS-1313490, DMS-1615288), the China Scholarship Council, the National Natural Science Foundation of China (Grant Nos.61379014) and the Natural Science Foundation of Tianjin ( No.16JCYBJC15900).

\ifCLASSOPTIONcaptionsoff
  \newpage
\fi



%

%

\begin{IEEEbiography}[{\includegraphics[width=1in,height=1.25in,clip,keepaspectratio]{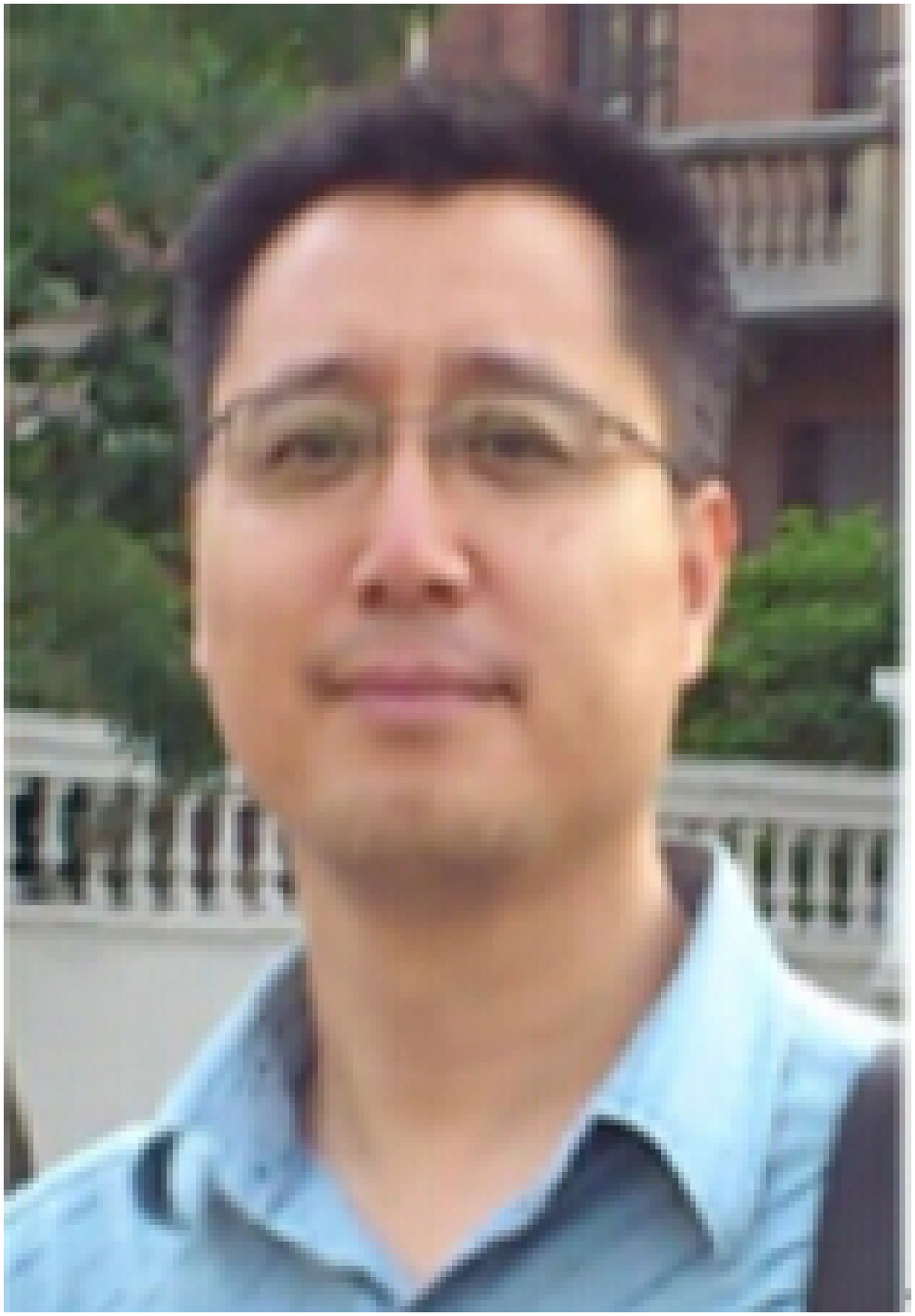}}]{Zhanjie Song}
received the Ph.D. degree in probability theory and mathematical statistics from the School of Mathematical Science, Nankai University, Tianjin, China. He is currently a Full Professor with the School of mathematics, a Research Affiliate with the State Key Laboratory of Hydraulic Engineering Simulation and Safety, and a Vice-Director with the Institute of TV and Image Information, all at Tianjin University (TJU), Tianjin. He was a Post-Doctoral Fellow in signal and information processing. His current research interests include approximation of deterministic signals, reconstruction of random signals, compressed sampling of multidimensional signals, and statistical analysis of random processes.
\end{IEEEbiography}

\begin{IEEEbiography}[{\includegraphics[width=1in,height=1.25in,clip,keepaspectratio]{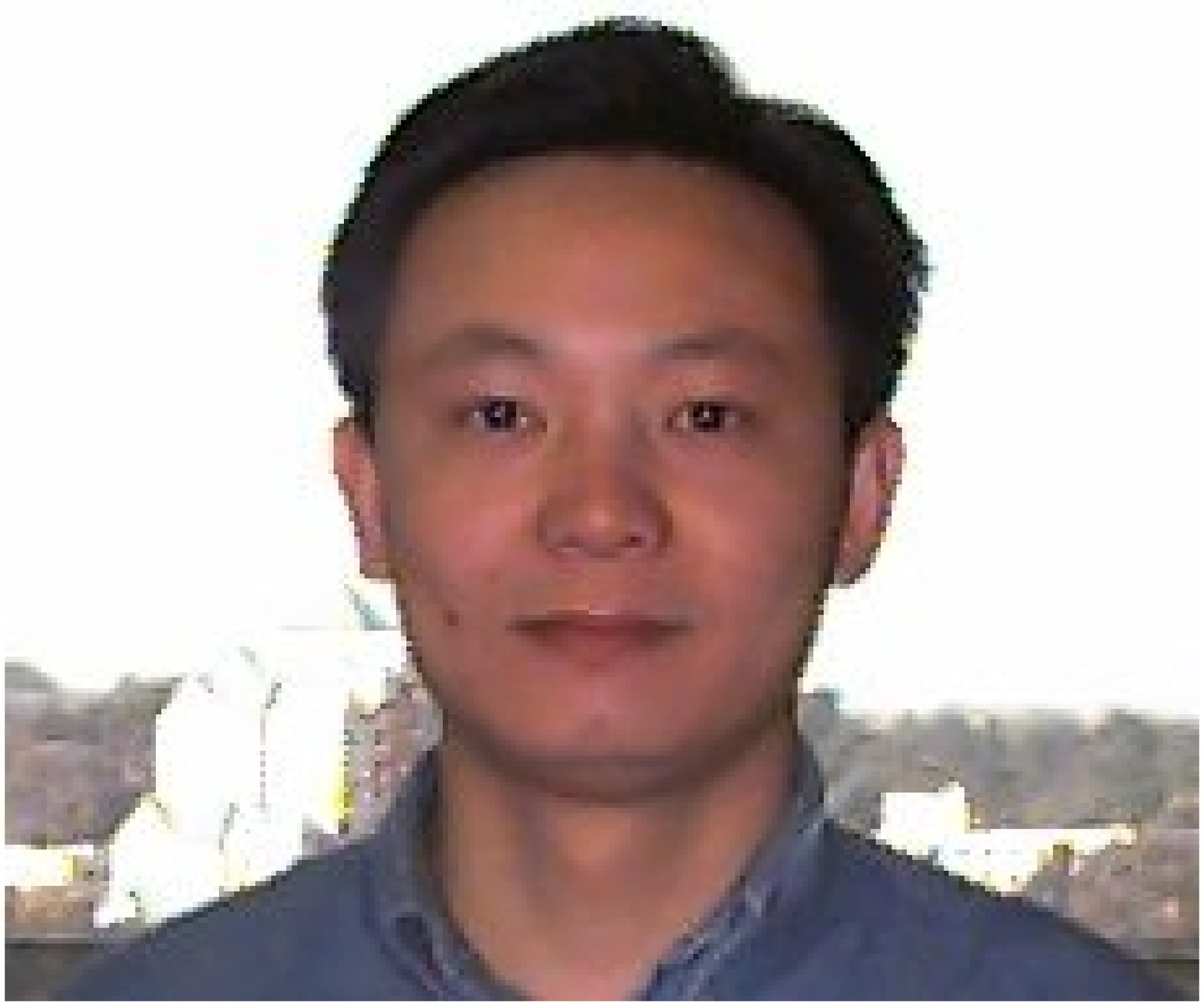}}]{Shidong Li}
received the M.S. degree in electrical engineering from the Graduate School of the Chinese Academy of Sciences, Beijing, China, the M.S. degree in applied mathematics from the University of Maryland, College Park, and the Ph.D. degree in applied mathematics from the Graduate School, University of Maryland, Baltimore, in 1985, 1989, and 1993, respectively. He was a Visiting Assistant Professor with Dartmouth College, Hanover, NH, from 1993 to 1994. From 1994 to 1996, he was with the University of Maryland. He joined San Francisco State University, San Francisco, CA, in 1996, where he has been a Tenured Full Professor of mathematics since 2005. His current research interests include frames and frame extensions and their applications in signal and image processing.

\end{IEEEbiography}

\begin{IEEEbiography}[{\includegraphics[width=1in,height=1.25in,clip,keepaspectratio]{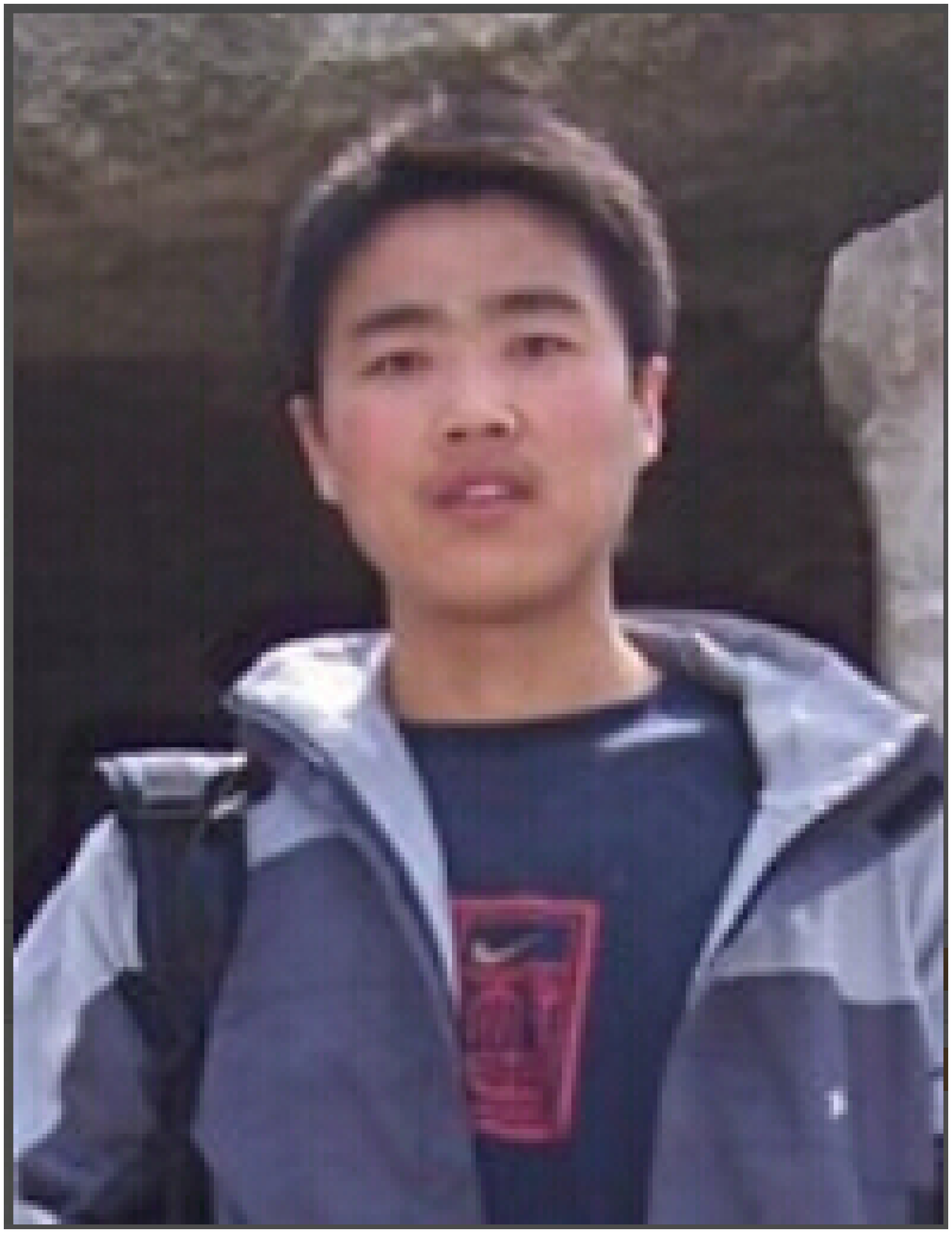}}]{Ningning Han}
 is currently pursuing the Ph.D. degree at Tianjin University. His work focused on the research of recovering the sparse or low-rank signal. Most recently, he has concentrated his efforts on direction of arrival (DOA) estimations working with Dr. Li. DOA is widely encountered in radar signal processing application platforms, extremely parallel with the classical harmonic retrieval problems.
\end{IEEEbiography}





\end{document}